\documentclass[manuscript,acmsmall,screen]{acmart}

\AtBeginDocument{%
  \providecommand\BibTeX{{%
    \normalfont B\kern-0.5em{\scshape i\kern-0.25em b}\kern-0.8em\TeX}}}

\usepackage[ruled,lined,linesnumbered,vlined]{algorithm2e}

\SetCommentSty{mycommfont}

\usepackage{amsmath}
\usepackage{amsfonts}
\usepackage{amsthm}
\usepackage{balance}
\usepackage{enumitem}
\usepackage[T1]{fontenc}
\usepackage{graphicx}
\usepackage{listings}
\usepackage{multicol}
\usepackage{multirow}
\usepackage{natbib}
\usepackage{caption}
\usepackage{subcaption}
\usepackage{url}
\usepackage{xspace}
\usepackage{float}
\usepackage{xcolor}
\usepackage{threeparttable}
\usepackage{tcolorbox}

\usepackage[normalem]{ulem}
\usepackage{longtable}
\usepackage{colortbl}
\usepackage{stfloats}
\usepackage{pifont}
\usepackage{fontawesome}
\usepackage{fancybox}

\usepackage{ulem}

\usepackage{booktabs} 

\newif\ifInCopypaper

\setcitestyle{numbers,square}

\usepackage{tikz}
\newcommand*\circled[1]{\tikz[baseline=(char.base)]{
        \node[shape=circle,draw,inner sep=0.5pt] (char) {\small#1};}}

\newcommand{\vs}{\hbox{\emph{vs.}}\xspace}

\newcommand{\eg}{\hbox{\emph{e.g.}}\xspace}
\newcommand{\ie}{\hbox{\emph{i.e.}}\xspace}

\newcommand{\etc}{\hbox{\emph{etc.}}\xspace}

\makeatother

\definecolor{mygreen}{HTML}{02818a}

\mathchardef\mhyphen="2D

\newcommand{\myparagraph}[1]{
  \vspace*{0.04cm}
  \noindent \textit{\textbf{#1.}}\quad
}
\newcommand{\mycode}[1]{\texttt{#1}\xspace}

\newcommand{\projectname}[1]{\textsf{#1}\xspace}

\newsavebox\mybox

\newcommand{\proj}{\projectname{LATTE}}

\newcommand{\piller}{P\xspace}
\newcommand{\class}{C\xspace}
\newcommand{\basic}{B\xspace}
\newcommand{\variant}{V\xspace}

\newcommand{\ghidra}{Ghidra\xspace}

\newcommand{\emtaint}{Emtaint\xspace}
\newcommand{\arbiter}{Arbiter\xspace}
\newcommand{\karonte}{Karonte\xspace}
\newcommand{\cwe}{CWE\xspace}
\newcommand{\cve}{CVE\xspace}

\newcommand{\newbugs}{37\xspace}
\newcommand{\cvebugs}{10\xspace}
\newcommand{\totaluniquebugs}{119\xspace}

\newcommand{\gptthree}{GPT-3.5\xspace}
\newcommand{\gptfour}{GPT-4.0\xspace}
\newcommand{\tps}{TPs\xspace}
\newcommand{\tp}{TP\xspace}
\newcommand{\fps}{FPs\xspace}
\newcommand{\fp}{FP\xspace}
\newcommand{\fnns}{FNs\xspace}
\newcommand{\fnn}{FN\xspace}

\newcommand{\tn}{TN\xspace}
\newcommand{\fscore}{F1\xspace}

\newcommand{\llm}{LLM\xspace}
\newcommand{\gpt}{GPT\xspace}
\newcommand{\llms}{LLMs\xspace}

\newcommand{\circledone}{\circled{1}\xspace}
\newcommand{\circledtwo}{\circled{2}\xspace}

\SetKwFunction{Algconstruct}{ConstructPromptSeq}
\SetKwFunction{Algdeduplicate}{Deduplicate}
\SetKwFunction{AlgExtractDynFuncNames}{ExtractExternalFunc}
\SetKwFunction{AlgBackwardslicing}{BackwardSlicing}

\SetKwData{AlgBinary}{DB}
\SetKwData{AlgPromptSeq}{PS}
\SetKwData{AlgPromptSeqs}{PSs}
\SetKwData{Algfuncext}{F$_\textit{external}$}
\SetKwData{Algssf}{sink}
\SetKwData{Algssfs}{sinks}
\SetKwData{AlgParam}{Para}
\newcommand{\Param}{Para}
\SetKwData{AlgLoc}{Loc}
\SetKwData{AlgVar}{Var}
\SetKwData{AlgArg}{Arg}

\SetKwData{AlgVulnerdest}{VD}
\SetKwData{AlgVulnerdests}{VDs}
\SetKwData{Algcallchain}{CC}
\SetKwData{Algcallchains}{CCs}
\SetKwData{Algextinpsource}{source}
\SetKwData{Algextinpsources}{sources}
\SetKwData{Algextinpsink}{sink}
\SetKwData{Algextinpsinks}{sinks}
\SetKwData{Algsource}{Source}
\SetKwData{Algcandidatedangflow}{CDF}
\SetKwData{Algcandidatedangflows}{CDFs}
\SetKwData{Algdangflow}{DF}
\SetKwData{Algdangflows}{DFs}
\SetKwData{Algprompttemp}{PT}
\SetKwData{Algprompttemps}{PTs}

\newcommand{\juliet}{Juliet\xspace}

\usepackage{tcolorbox}
\newcounter{rqnumber}
\newcommand{\finding}[1]{
    \begin{tcolorbox}
        \textbf{Answer to RQ\refstepcounter{rqnumber}\therqnumber}: #1
    \end{tcolorbox}
}

\hypersetup{hidelinks}

\usepackage{cleveref} 
\Crefname{table}{Table}{Tables}
\crefname{table}{Table}{Tables}
\Crefname{figure}{Fig.}{Figures}
\crefname{figure}{Fig.}{Figures}
\Crefname{algocf}{Algorithm}{Algorithms}
\crefname{algocf}{Algorithm}{Algorithms}
\Crefname{algorithm}{Algorithm}{Algorithms}
\crefname{algorithm}{Algorithm}{Algorithms}
\crefname{appendix}{Appendix}{Appendices}
\Crefname{appendix}{Appendix}{Appendices}

\crefformat{chapter}{\S~#2#1#3}
\crefmultiformat{chapter}{\S\S~#2#1#3}{ and~#2#1#3}{, #2#1#3}{, and~#2#1#3}

\crefformat{section}{\S~#2#1#3}
\crefmultiformat{section}{\S\S~#2#1#3}{ and~#2#1#3}{, #2#1#3}{, and~#2#1#3}

\setcopyright{acmlicensed}
\acmJournal{TOSEM}
\acmYear{2025} \acmVolume{1} \acmNumber{1} \acmArticle{1} \acmMonth{1}\acmDOI{10.1145/3711816}

\begin{document}

\title{Harnessing the Power of LLM to Support Binary Taint Analysis}

\author{Puzhuo Liu}
\affiliation{%
\institution{Ant Group; Tsinghua University}
\country{China}
}
\email{liupuzhuo.lpz@antgroup.com}

\author{Chengnian Sun}
\affiliation{%
  \institution{University of Waterloo}
  \city{Waterloo}
  \state{Ontario}
  \country{Canada}
}
\email{cnsun@uwaterloo.ca}

\author{Yaowen Zheng }
\affiliation{%
  \institution{Institute of Information Engineering, CAS}
  \state{Beijing}
  \country{China}}
\email{zhengyaowen@iie.ac.cn}

\author{Xuan Feng}
\affiliation{%
  \institution{Independent Researcher}
  \city{Vancouver}
  \state{British Columbia}
  \country{Canada}
  }
\email{xfeng9209@gmail.com}

\author{Chuan Qin}
\affiliation{%
\institution{Institute of Information Engineering, CAS; University of Chinese Academy of Sciences}
  \state{Beijing}
  \country{China}
}
\email{qinchuan@iie.ac.cn}

\author{Yuncheng Wang}
\affiliation{%
\institution{Institute of Information Engineering, CAS; University of Chinese Academy of Sciences}
  \state{Beijing}
  \country{China}
}
\email{wangyuncheng@iie.ac.cn}

\author{Zhenyang Xu}
\email{zhenyang.xu@uwaterloo.ca}
\affiliation{
    \institution{University of Waterloo}
    \city{Waterloo}
    \state{ON}
    \country{Canada}
    \postcode{N2L 3G1}
}

\author{Zhi Li}
\affiliation{%
\institution{Institute of Information Engineering, CAS; University of Chinese Academy of Sciences}
  \state{Beijing}
  \country{China}
}
\email{lizhi@iie.ac.cn}

\author{Peng Di}
\authornote{Corresponding Author}
\affiliation{%
\institution{Ant Group}
\state{Zhejiang}
\country{China}
}
\email{dipeng.dp@antgroup.com}

\author{Yu Jiang}
\affiliation{%
\institution{Tsinghua University}
\state{Beijing}
\country{China}
}
\email{jy1989@mail.tsinghua.edu.cn}

\author{Limin Sun}
\affiliation{%
\institution{Institute of Information Engineering, CAS; University of Chinese Academy of Sciences}
  \state{Beijing}
  \country{China}
}
\email{sunlimin@iie.ac.cn}

\renewcommand{\shortauthors}{Puzhuo Liu, et al.}

\begin{abstract}
This paper proposes \proj,
the first static binary taint analysis
that is powered by a large language model (\llm).
\proj is superior to the state of the art
(\eg, \emtaint, \arbiter, \karonte)
in three aspects.
First, \proj is fully automated while
prior static binary taint analyzers need
rely on human expertise to manually customize taint propagation rules and vulnerability inspection rules.
Second, \proj is significantly effective in
vulnerability detection,
demonstrated by our comprehensive evaluations.
For example,
\proj has found \newbugs new bugs in
real-world firmware,
which the baselines failed to find.
Moreover, \cvebugs of them have been assigned \cve numbers.
Lastly, \proj incurs remarkably low engineering cost,
making it a cost-efficient and scalable solution for security researchers and practitioners.
We strongly believe that \proj opens up a new direction
to harness the recent advance in \llms to improve
vulnerability analysis for binary programs.

\end{abstract}



\maketitle

\section{Introduction}

For decades,
software testing has remained the dominant
method for software quality assurance, primarily due to its cost-effectiveness.
However, as Edsger Dijkstra emphasized~\cite{dijkstra2002ewd},
software testing can only reveal the presence of bugs, not
their absence.
Consequently, it is not uncommon to encounter bugs in released binaries~\cite{vadayath2022arbiter,7546500}.
Certain bugs can be particularly critical and may be exploited as vulnerabilities
to initiate cyberattacks,
\eg, unauthorized access to systems, stealing sensitive data, and disrupting normal system operations~\cite{miria,Stuxnet}.

To reduce the impact of vulnerabilities in released binaries
such as firmware of embedded devices~\cite{DECAF,redini2020karonte,zheng2019firm,cheng2023emtaint}
and commercial software applications~\cite{busch2023teezz,QueryX,pewny2015cross,vadayath2022arbiter},
third-party security audits emerged as an indispensable force.
These audits,
conducted by entities without access to the source code,
apply various post-development
analysis~\cite{DECAF,qiao2017function,meng2016binary,zhang2021osprey}
and testing techniques
~\cite{redini2020karonte,zheng2019firm,cheng2023emtaint,vadayath2022arbiter}
on the binaries.
\emph{Thus, though analysis and testing are usually applied on source code,
analyzing and testing binaries is also an indispensable, complementary technique
to ensure software reliability and security}.

\myparagraph{Static Binary Taint Analysis}
Among these analysis and testing techniques~\cite{zheng2019firm,cheng2023emtaint,DECAF},
static binary taint analysis (referred to as \emph{taint analysis} for brevity in this paper)
stands out as an effective approach for vulnerability inspection.
It can directly analyze binaries with several advantages, including
high test code coverage,
independence from the concrete execution environment,
and the ability to discover previously unknown exploitable
vulnerabilities~\cite{vadayath2022arbiter,cheng2023emtaint,redini2020karonte}.

The standard taint analysis workflow comprises three steps.
\circledone \textit{Identifying Taint Sources}:
Taint sources are typically functions that receive external data.
This external data can be manipulated by attackers.
For example, the function \mycode{recv} reads data through a socket,
\mycode{getenv} retrieves information from an environment variable,
and \mycode{fgets} loads data from a file.
\circled{2} \textit{Propagating Taint Labels}:
Initially, the external data received through taint sources
are labeled tainted.
Then the analysis traces the flow of the tainted data through the program;
it propagates the taint labels along data dependencies, such as
assignments, computations, and function calls,
to discover how the tainted data spreads throughout the program.
Meanwhile, if a memory region with a taint label is overwritten with untainted data
or the region is sanitized with certain checks,
then the taint label is removed from that memory region.
\circled{3} \textit{Identifying and Inspecting Sinks}:
Sinks are potential points in the program where tainted data might
be misused or lead to security vulnerabilities.
They are usually sensitive operations including
manipulating strings (\eg, \mycode{strcpy}, \mycode{strcat}),
executing commands (\eg, \mycode{system}, \mycode{execl}),
and formatted outputting (\eg, \mycode{printf}, \mycode{fprintf}).

\myparagraph{Limitations of Taint Analysis}
Taint analysis remains a partially automated process,
demanding considerable human effort for its execution.
The effectiveness of this analysis heavily relies on
the accuracy and quality of rules
governing taint source, taint label propagation and sink inspection.
In particular, using taint analysis for vulnerability detection
not only involves the effective propagation of taints,
but also requires setting up as complete taint checking rules as possible to reduce
false positives and false negatives.
Regrettably, designing these rules necessitates
meticulous manual crafting,
resulting in a non-trivial and error-prone task.
Previous techniques~\cite{redini2020karonte,cheng2023emtaint,vadayath2022arbiter,liu2022finding,gotovchits2018saluki,cheng2018dtaint}
have relied on human expertise to manually define these rules,
thereby introducing the risk of false positives or false negatives
in the analysis results due to potential rule errors and absences~\cite{liu2022finding,chen2021sharing}.
For instance, determining taint sources involves identifying functions that receive external data,
and establishing the initial taint labels requires the specification of these functions.
Additionally, the sanitization checks that transform tainted data
into safe data in propagation rules
are diverse and specific to function semantics.
Furthermore, the inspection rules also vary based on the types of
vulnerabilities and sinks involved (further details in \cref{sec:Motivation}).

\myparagraph{Insights}
The limitation of taint analysis primarily arises from the requirement of
an in-depth understanding of the semantics of the code and the vulnerability pattern.
Extension of rules has historically imposed a significant burden on humans
when applying taint analysis to different binaries or vulnerabilities~\cite{cheng2018dtaint,meng2016binary,vadayath2022arbiter}.
To overcome this limitation, we firmly believe that large language models (\llms)
can provide valuable assistance,
benefiting from their recent advances in code understanding,
including code generation~\cite{yeticstiren2023evaluating,perry2022users,xia2023conversational}
and code summarization~\cite{ahmed2022few,sun2023automatic}.
\llms have advantages in code-related tasks
by providing
context awareness,
semantic understanding, and
natural language interaction.
While \llms have their advantages, it is essential to note that they are not a silver bullet.
Therefore, in this paper,
we embark on the first study for vulnerability discovery
that explores the synergistic combination of
the ability of \llm in code understanding
and the concept of taint analysis.
This will improve the automation and accuracy of taint-style vulnerability inspections.

\myparagraph{Challenge of Using \llms}
There are unexplored challenges
in applying \llms for vulnerability inspection in binaries.
\circledone Prompt engineering:
Prompts play a crucial role in instructing \llms
to generate responses, and their design significantly influences the performance of \llms.
In the context of vulnerability analysis,
well-constructed prompts must include
the code snippets to be analyzed
and a clear description of the analysis tasks.
However, current \llms have their limitations:
providing all the code snippets related to potential
vulnerabilities at once or directly instructing vulnerability
analysis often leads to unsatisfactory results.
Therefore, it is preferable to
break down the relevant code snippets and vulnerability analysis tasks,
and prompt \llms to perform step-by-step analysis.
Furthermore, the prompt engineering process should be
designed to be general, without requiring human intervention
whenever the program or vulnerability analysis task changes.
\circledtwo Encoding of code context:
To conduct accurate vulnerability analysis,
having adequate code context, including function call stacks, data dependencies,
and control flow, is crucial to determine
whether a piece of code is vulnerable.
However, most \llms currently accept only a limited amount of context,
striking a balance between computational costs and response quality.
For example, the state-of-the-art \llm GPT-4.0 supports contexts
consisting of just 32,000 tokens (approximately 24,000 words) for both prompts and responses~\cite{token}.
Consequently, when dealing with real binaries,
encoding the code context to meet this size constraint of \llms
becomes necessary.
Analyzing the entire binary directly is often impractical, as the decompiled code
itself usually surpasses this size limitation.

\myparagraph{\proj}
We propose \proj
(\underline{L}LM-Powered Bin\underline{a}ry \underline{T}ain\underline{t} Analyz\underline{e}r),
the first technique to inspect vulnerabilities in released binaries
by combining the power of \llms
and the concept of taint analysis.
It conquers the practical limitations of taint analysis
by harnessing the ability of \llms in code understanding to automate and streamline the manual components of taint analysis,
and addresses the challenges of using \llms with dedicated static analysis and prompt engineering.
The key idea of \proj is to construct prompt sequences driven by code slicing.
Concretely,
\begin{enumerate}[leftmargin=*,topsep=0em]
\item 	Initially,
we preprocess the binary
to recover the code structure,
by performing analysis such as disassembly and decompilation.

\item Based on the preprocessing results,
we use \llms to identify sinks and then
determine potentially vulnerable destinations, \ie,
call sites of security sensitive functions (sinks).
These vulnerable destinations will need to be inspected.

\item
With the identified vulnerable destinations,
we
perform \emph{backward} intraprocedural and interprocedural data dependency slicing
to establish function call chains starting from the vulnerable destinations.

\item
We consider function call chains involving external input data as \emph{dangerous flows}.
To this end, we first use \llms to identify sources that pass externally controllable data into the program
and then identify such call chains that include sources.

\item
Based on the dangerous flows,
we use functions as units and
combine them with program analysis tasks, such as flow analysis,
alias analysis and taint analysis, to form prompt sequences.
These prompt sequences are generated to converse with \llms,
completing the vulnerability inspection.
\end{enumerate}

\myparagraph{Evaluations}
Our comprehensive evaluations demonstrate that \proj significantly outperforms
the state-of-the-art binary taint analyzers.
Specifically, We first evaluated \proj
based on the taint-style vulnerability test subjects
(compiled and stripped) in the standard dataset Juliet Test Suite~\cite{juliet}.
The accuracy and F1 of \proj vulnerability checking
exceeds \emtaint~\cite{cheng2023emtaint} and \arbiter~\cite{vadayath2022arbiter} on every vulnerability type.
Moreover,
\proj achieved 100\% correct coverage for identifying sinks
and sources in the test subject target scenario.
For test subjects labeled with containing vulnerabilities,
\proj successfully extracted more than 95\%
of the correct dangerous flows
that covered the paths triggering potential vulnerabilities.
Furthermore, we assessed the effectiveness of \proj
in real-world binaries using the embedded device firmware dataset
provided by \karonte~\cite{redini2020karonte}.
\proj detected a total of \totaluniquebugs unique bugs,
including \newbugs previously unknown bugs (\cvebugs CVE numbers have been given due to high threat),
which outperforms the state of the art~\cite{redini2020karonte,cheng2023emtaint,vadayath2022arbiter}.

\myparagraph{Contributions}
We make the following major contributions.

\begin{itemize}[leftmargin=10pt,topsep=0pt]

\item We propose \proj, the first technique to leverage \llms to power static binary taint analysis.
\proj addresses the limitations of prior taint analysis.
Prior taint analyzers heavily rely on human expertise to define rules for
propagating taint labels and inspecting sinks and customize rules for different binaries and vulnerability types,
whereas \proj harnesses the ability
of \llms in code understanding to fully automate taint analysis.

\item \proj presents a novel solution to address the
practical challenges of using \llms, with
dedicated multiple-step static analyses
and automated prompt engineering
to effectively elicit
\llms to generate accurate analysis responses.

\item Comprehensive evaluations demonstrate
the effectiveness of \proj.
Compared to the state of the art,
\proj achieved higher accuracy and F1 score on the standard dataset,
and on the firmware dataset, \proj found 117 unique bugs (including 37 previously unknown bugs
with \cvebugs awarded \cve numbers),
at least 21 more than the baselines.

\end{itemize}

\section{Background}

\subsection{Taint-Style Vulnerability}\label{subsec:vulnerability-patterm}

As an effective software testing technique,
taint analysis is often used for binary vulnerability
detection~~\cite{cheng2018dtaint,cheng2023emtaint,gotovchits2018saluki,redini2020karonte,liu2022finding,chen2021sharing,redini2017bootstomp}.
Taint analysis can effectively find
vulnerabilities caused by external inputs,
commonly referred to as taint-style vulnerabilities~\cite{yamaguchi2015automatic}.
Nevertheless,  previous binary taint techniques~\cite{cheng2018dtaint,cheng2023emtaint,gotovchits2018saluki,redini2020karonte,liu2022finding,chen2021sharing,redini2017bootstomp}
use different names to denote types of vulnerabilities they support.
To facilitate comparison of different techniques in a unified, standard way,
we categorize the vulnerability types
supported by previous binary taint analyzers according to the
Common Weakness Enumeration (\cwe)~\cite{cwe}.

\begin{figure}[t]
	\centering
	\includegraphics[width=0.7\linewidth]{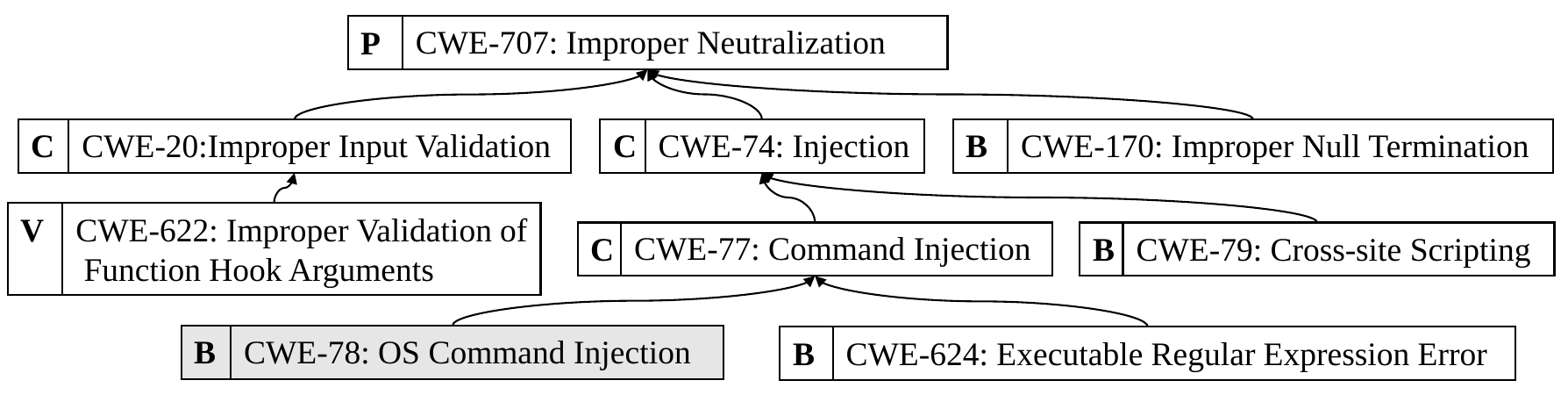}
	\caption{Example of the class CWE-78.}
	\label{fig_cwe_os}
\end{figure}

\cwe is a community-developed list of software weakness types
used to classify security vulnerabilities~\cite{cve}.
\cwe is a tree-like relationship between weaknesses at different levels of abstraction, which can be divided into \textit{pillars}, \textit{classes}, \textit{bases}, and \textit{variants}.
As shown in \cref{fig_cwe_os}, the \textit{pillar} (\piller) is the weaknesses described in the most abstract way.
The \textit{class} (\class) is still  abstract and usually independent of any particular language or technology.
The \textit{basic} (\basic) is used to present more specific types of weaknesses.
The \textit{variant} (\variant) is described at a  low level of details, usually restricted to a specific language or technology.

To ensure that vulnerability types do not overlap,
we classify taint-style vulnerability types
according to the following principles.
\circled{1} Vulnerability types are mapped to the \cwe's basic level,
which is used to present specific types of weaknesses.
If basic level does not exist, then the variant level is used.
\circled{2} Classification is based on the cause the vulnerability is triggered,
regardless of the reason for the introduction of tainted data.
Otherwise a vulnerability would have multiple labels.
For example, if a command injection vulnerability (CWE-78)
is caused by unchecked input data, it will be classified as CWE-78,
not CWE-76 (Improper Neutralization of Equivalent Special Elements).
\circled{3} Classification is based on the first reason the vulnerability is triggered.
For instance, if there is an integer overflow vulnerability (CWE-190)
in the operation of input data,
and the subsequent use of the operation result leads
to a buffer overflow (CWE-680: Integer Overflow to Buffer Overflow),
we still classify it as CWE-190.

\begin{table*}[h]
	\caption{Mapping between taint-style vulnerability types and CWE, comparison of binary taint analyzer capabilities.
		``\checkmark{}'' means that the corresponding taint analyzer supports the corresponding vulnerability type.
		``*'' indicates that the taint analyzer \arbiter does not provide the corresponding inspection rules and additional engineering effort is required.
	}
	\label{tab:cweandtainter}
	\resizebox{\textwidth}{!}{%
		\begin{tabular}{@{}llcccccccc@{}}
			\hline
			\multicolumn{2}{l}{\multirow{2}{*}{\textbf{Vulnerability Type} } }                       & \multicolumn{8}{c}{\textbf{Binary Taint Analyzer}}                                \\ \cline{3-10}
			&                               & Saluki~\cite{gotovchits2018saluki} & Dtaint~\cite{cheng2018dtaint} & Karonte~\cite{redini2020karonte} & SATC~\cite{chen2021sharing} & FBI~\cite{liu2022finding} & Arbiter~\cite{vadayath2022arbiter} &  Emtaint~\cite{cheng2023emtaint}  & \proj \\ \hline
			CWE-78 &OS Command Injection           & \checkmark                      &    \checkmark               &                                   &    \checkmark              &   \checkmark                &   \checkmark $^*$           &   \checkmark       &    \checkmark  \\
			CWE-120 &Classic Buffer Overflow          &                                  &   \checkmark           &   \checkmark                         &      \checkmark          &    \checkmark                  &   \checkmark$^*$               &  \checkmark         &    \checkmark  \\
			CWE-134& Controlled Format String         &                                  &                        &                                     &                   &   \checkmark                              &   \checkmark         &   \checkmark  &    \checkmark  \\
			CWE-190 &Integer Overflow or Wraparound &                                     &                       &                                   &                              &                                        &   \checkmark       &           &   \checkmark   \\
			CWE-606& Unchecked Loop Condition   &                                          &                      &   \checkmark                      &                          &                                     &                            &           &   \checkmark   \\ \hline
		\end{tabular}%
	}
\end{table*}

\cref{tab:cweandtainter}
lists five \cwe vulnerability types,
which cover the vulnerability types 
supported by the previous binary taint analyzer.
Previous techniques rely on
human expertise or heuristics to define taint analysis rules
including taint sources, sanitizers, and sinks,
to propagate and inspect tainted data
to discover taint-style vulnerabilities in binaries.
While theoretically it is possible to apply a well-implemented taint engine
and sound rules to inspect all types of vulnerabilities in any binary, this is not realistic.\footnote{A taint engine is orthogonal to taint rules.
A taint engine is the tool that actually performs taint analysis,
detecting potential security vulnerabilities by tracking data flows.
Taint rules are a set of principles and logic that guide the work of the taint engine,
defining which data is tainted  and how to track and process this data.
}

First, implementing a taint engine demands substantial engineering effort,
particularly for analyzing binary code.
Binaries often lack information on symbols and data types,
and can vary significantly due to differences in architecture,
compilers, and optimization options.
Even in decompiled code,
much semantic and structural information is lost or ambiguous.
Consequently, these previous techniques
typically have to make certain assumptions
to develop new taint
engines~\cite{redini2020karonte,cheng2023emtaint,gotovchits2018saluki,vadayath2022arbiter,cheng2018dtaint}.

Second,
defining sound taint source,
sanitization, and sink rules
for existing taint engines is
also a complex engineering task.
Methods such as function abstraction and machine learning
can assist in setting relevant rules for new vulnerability types
and libraries~\cite{liufits,chen2021sharing,liu2022finding}.
However, to improve the accuracy of vulnerability analysis
and reduce the workload of post-analysis,
the complexity of the rules
and the demand for automatically verifying vulnerabilities inevitably increase.
For example,
to reduce false positives,
certain taint engines~\cite{redini2020karonte,cheng2023emtaint} perform vulnerability inspections
at the sinks of \mycode{strcpy}, \mycode{strncpy}, and \mycode{memcpy}
by comparing the length of the taint source and the size of
the destination buffer;
these rules increase the analysis complexity and workload,
and demand more insights from expert experience.

\subsection{Large Language Model}
An \llm consists of billions of parameters and is trained on billions of samples to be highly general.
Downstream tasks can be accomplished using \llms through either prompt engineering~\cite{lemieux2023codamosa,deng2023large,kang2023large,xia2023keep}
or fine-tuning~\cite{deng2023fuzzgpt}.
In the context of prompt engineering,
the downstream task is directly provided as a natural-language description to the \llm,
prompting the \llm to generate a response to complete the task.
Unlike fine-tuning,
which involves retraining the \llm with additional data,
prompt engineering does not necessitate extra training data
and avoids the high computational resource costs.
Moreover,
the \llm has been trained on extensive corpora
and possess the ability
to execute specific tasks with prompt engineering
~\cite{bubeck2023sparks,chatafl,li2023assisting,guo2024exploring,xia2023keep}.
Therefore, we chose the prompt engineering method to conduct our research.

Prompt engineering is a critical process that involves constructing prompts
to elicit desired responses from \llms.
It plays a direct role in influencing the performance of \llms across various tasks.
With deliberated prompt engineering, \llms
have demonstrated excellent code understanding capabilities on many code-related tasks,
such as code generation~\cite{yeticstiren2023evaluating,perry2022users,xia2023conversational}
and code summarization~\cite{ahmed2022few,sun2023automatic}.
These successes inspired
us to explore extending the power of \llms into the field of vulnerability analysis.

\section{Motivation}\label{sec:Motivation}
Benefiting from the rich code information 
and community resources, 
the development of source code vulnerability static analysis tools 
has far surpassed that of binary static analysis tools~\cite{codeql,coverity,infer}.
Compared with source code, in binary, even decompiled pseudo code, 
there is still a lack of clear functions, classes, variable names and comments, 
which makes it difficult to understand specific functions.
\cref{tab:cweandtainter}
demonstrates that each prior binary taint technique only supports
a limited set of taint-style vulnerability types.
This limitation arises from the difficulty in manually defining comprehensive propagation and inspection rules for taint analysis.
Efforts such as SATC~\cite{chen2021sharing} and FBI~\cite{liu2022finding} have attempted to optimize
Karonte's~\cite{redini2020karonte} rules using human expertise and heuristics,
resulting in reduced false negatives and false positives.
However, due to the complexity and variety of taint analysis,
the manual-dependent nature still hinders 
the overall performance of these binary taint analyzers.
As shown in \cref{fig:integer_example},
we take the integer overflow vulnerability
in a decompiled binary program as an example,
to intuitively illustrate the following concrete 
problems of designing and implementing
effective binary taint analyzer.

\begin{figure}[t]
    \begin{minipage}{0.8\linewidth}
    \begin{lstlisting}[frame=leftline,xleftmargin=1em, numbers=left,
      numbersep=5pt,showstringspaces=false,basicstyle=\footnotesize\ttfamily,escapechar=|]
void |\textbf{foo}|(void) {
  signed char a, b, c, d;
  int e;
  fscanf(stdin, "%d", &a); |\label{line:fscanf}|
  b = a; |\label{line:prop1}|
  c = 0;
  a = 1; |\label{line:san}|
  e = b + 1; |\label{line:add}| //No integer overflow occurs after type conversion
  printf("%d",(ulong)e); |\label{line:print1}|
  if (b == 127){  |\label{line:if}|
    printLine("Data value is too large");
  }
  else{
    c = b + 1;  |\label{line:add2}| //No integer overflow occurs after sanitization
    printf("Result is %d\n", c); |\label{line:print2}|
  }
  d = c + 1; |\label{line:add3}| //The input is 126 causing an integer overflow
  printf("Result is %d\n", d);
  return;
}
    \end{lstlisting}
    \end{minipage}
\caption{
Example of vulnerabilities in decompiled code (comments are added to aid understanding).
When the user input from stdin is 126, the expected result on line 17 is 128 (the actual result is -128). Because the char type is signed, its range is [-128, 127]. When the result exceeds this range, an integer overflow occurs.}
\label{fig:integer_example}
\end{figure}

\newcommand{\myreason}[1]{
\vspace*{1pt}
\noindent\uline{#1.}\quad
}

\subsection{Problem 1. Determining Taint Sources and Initial Taint Labels}
Manually identifying taint sources, which receive external input
in a binary, is a laborious process,
because not only standard C/C++ functions
(such as \mycode{recv}, \mycode{fscanf}, \mycode{fgets})
but also third-party functions (\eg, \mycode{SSL\_read} and \mycode{BIO\_read} in OpenSSL)
can receive external data.
More importantly, determining the initial taint labels
for the return values and parameters
of the taint sources
requires deep understanding of the semantics of the taint sources.
For example,
only with the semantics of \mycode{fscanf}
can we determine
the \mycode{third} parameter \mycode{\&a} of \mycode{fscanf} (\cref{line:fscanf})
to be tainted, as \mycode{\&a} stores the external input retrieved by \mycode{fscanf};
the first and second parameters should not be labeled as tainted.
Note that it is even more challenging to determine initial taint labels for
third-party functions than standard C/C++ functions because the former
usually do not have  documentation whereas the latter are at least documented
in the C/C++ language standard.
Mislabeling often leads to the incorrect propagation
of taint information throughout the program under analysis.

\subsection{Problem 2: Defining Rules to Propagate Taint Labels}
Initial taint labels are propagated throughout the program
with assignments (\eg, \mycode{b=a} on \cref{line:prop1}),
calculations (\eg, \mycode{e=b+1} on \cref{line:add}) and other language constructs,
which can be automatically identified through data dependency analysis.
The process of propagating taint labels involves not only
propagation
but also
sanitization.
In addition to the sanitization performed by directly assigning or copying safe (\ie, untainted)
data to tainted memory region (\eg, \mycode{a=1} on \cref{line:san}),
there are also sanitization situations related to semantics.
For example,
checking the content of \mycode{b} on \cref{line:if}
excludes the possibility of overflow of \mycode{c=b+1} on \cref{line:add2}.
Due to the considerable engineering effort caused by the diversity of sanitization situations
and the correlation of code semantics,
previous work~\cite{redini2020karonte,liu2022finding,cheng2023emtaint}
ignores the setting of complex sanitization,
which leads to
considerable false positives, as discussed in~\cite{cheng2023emtaint,10.1145/1543135.1542486,marashdih2023enhanced}.

\subsection{Problem 3: Formulating Inspection Rules for Various Sinks and Vulnerabilities}
Taint analysis can discover data flow sensitive vulnerabilities caused by external inputs,
including various types of vulnerabilities listed in \cref{tab:cweandtainter}.
These vulnerabilities can be triggered by different sinks in various ways,
leading to a complex many-to-many
relationship between vulnerability types and sinks.
For each pair of a vulnerability type and a sink, it is often necessary
to manually formulate a customized inspection rule.
The process starts with identifying sinks, which can include not only standard C/C++ functions
like \mycode{printf} and \mycode{system} but also third-party functions from various libraries
like \mycode{BIO\_printf} in OpenSSL~\cite{openssl}.
When taint analysis is applied to a new binary that is linked with different libraries,
human effort is indispensable to identify the sinks,
because sink identification
involves comprehending the behavior of functions in the binary.
After determining sinks, inspection rules need to be formulated according to the potential vulnerability types.

For example, to inspect whether the \mycode{printf} call on \cref{line:print1}
in \cref{fig:integer_example} can trigger a CWE-134 vulnerability,
we need to check whether the first argument (\ie, \mycode{"\%d"}) of \mycode{printf} is tainted;
to inspect whether
the call can trigger a CWE-190 vulnerability based on the first argument \mycode{\%d},
we first need to check whether the second argument (\ie, \mycode{(ulong)e}) of \mycode{printf} is tainted;
if yes,
we need to further check whether the calculation process of this argument overflows.
It is not scalable or practical to depend solely on human expertise and efforts
to understand sink functions and vulnerability types,
and manually formulate accurate and comprehensive
inspection rules~\cite{cheng2023emtaint,vadayath2022arbiter,redini2020karonte}.

\subsection{Advantages of \llms in Vulnerability Inspection}
The aforementioned specific implementation of the taint analysis
often requires manual efforts combined together with human expertise of vulnerability analysis
to establish taint rules.
Comprehensive taint rules can reduce the workload of post-processing analysis results,
but also increase the complexity of the rules and the cost of analysis.
For example,
if the length of the tainted string with the null-terminator
meets the requirements of the destination buffer of \mycode{strcpy},
the tainted string will not cause a buffer overflow vulnerability.
Traditional taint analysis tracks the flow of taint labels from sources to sinks,
so post-processing to verify alerts is critical, such as bounds checking and value analysis.

Fortunately, \llms inherently have the ability of human-like semantic understanding
of code,
unlike static or automated tools that rely on predefined patterns and rules.
Certain software engineering tasks,
such as code generation based on natural language descriptions~\cite{yeticstiren2023evaluating,perry2022users,xia2023conversational}
and code summarization ~\cite{ahmed2022few,sun2023automatic},
have demonstrated the superb ability of \llms to understand code.

To intuitively illustrate the benefits of LLM-assisted vulnerability inspection,
we show the vulnerability analysis results produced by GPT-4.0 for the following three code snippets
with this prompt template:
\begin{quote}\it
	Use \mycode{<function>} as the taint source
	and mark the \mycode{<parameter>} parameter with the taint label for taint propagation.
	Determine whether the program has vulnerabilities.
	Pay attention to the buffer size and library function definition.
	If there is a vulnerability, please explain it according to the CWE type.
\end{quote}
where
the pair (\mycode{<function>}, \mycode{<parameter>})
is replaced with
(\mycode{fgets}, \mycode{first})
or
(\mycode{recv}, \mycode{second}) accordingly.

\myparagraph{Code Snippet without Vulnerabilities}
The program in
\cref{fig:gpt-no-vulnerability}
 does not have any vulnerability and \gpt 4 correctly
produces the following analysis result.
\begin{quote}\it
    In the provided code snippet,
    using \mycode{fgets} as the taint source and marking
    the first parameter with the taint label for taint propagation,
    the primary concern revolves around how data is handled and copied between buffers.
    However, the specific use of \mycode{fgets} and \mycode{strcpy} in this context
    does not inherently introduce a classic buffer overflow vulnerability
    as might be seen with other functions like \mycode{gets} or unchecked uses of \mycode{strcpy}.
\end{quote}

\begin{figure}[h]
    \centering
    \begin{minipage}{0.8\textwidth}
    \begin{lstlisting}[frame=leftline,xleftmargin=1em, numbers=left,numbersep=5pt,showstringspaces=false,basicstyle=\footnotesize\ttfamily,escapechar=|]
int |\textbf{main}|() {
    char a[50];
    char b[50];
    printf("Enter a string: ");
    fgets(a, sizeof(a), stdin);
    strcpy(b, a);
    return 0;
}
    \end{lstlisting}
    \end{minipage}
    \vspace*{-1em}
    \caption{A decompiled binary program that has no vulnerability.
    }
    \label{fig:gpt-no-vulnerability}
    \end{figure}

\myparagraph{Vulnerable Code Snippet with \mycode{fgets}}
The program in
\cref{fig:gpt-with-fgets-vulnerability}
 is similar to
\cref{fig:gpt-no-vulnerability}
, but
has a buffer overflow vulnerability.
The following shows the correct analysis result by \gpt 4.
\begin{quote}\it
    In this updated code snippet, there is a clear vulnerability due to the use of
    \mycode{strcpy} to copy data from a larger buffer \mycode{a} (60 bytes)
    to a smaller buffer \mycode{b} (50 bytes) without checking the size of the input.
    This scenario introduces a classic buffer overflow vulnerability,
    which can be classified under CWE-120:
    Buffer Copy without Checking Size of Input (Classic Buffer Overflow).
\end{quote}

\begin{figure}[h]
    \centering
    \begin{minipage}{0.8\textwidth}
    \begin{lstlisting}[frame=leftline,xleftmargin=1em, numbers=left,numbersep=5pt,showstringspaces=false,basicstyle=\footnotesize\ttfamily,escapechar=|]
int |\textbf{main}|() {
    char a[60]; // The size of the buffer changes
    char b[50];
    printf("Enter a string: ");
    fgets(a, sizeof(a), stdin);
    strcpy(b, a); // A buffer overflow vulnerability here.
    return 0;
}
    \end{lstlisting}
    \end{minipage}
    \vspace*{-1em}
    \caption{A decompiled binary program, similar to  \cref{fig:gpt-no-vulnerability}, uses \mycode{fgets} to receive user input (comments are added to aid understanding).
        It has a buffer overflow vulnerability.
        The reason for this program to have a buffer overflow vulnerability is that
        the buffer that \mycode{fgets} receives input is 60 bytes long,
        and \mycode{strcpy} may exceed the buffer \mycode{b} of the destination when copying.
    }
    \label{fig:gpt-with-fgets-vulnerability}
    \end{figure}

\myparagraph{Vulnerable Code Snippet with \mycode{recv}}
The program in
\cref{fig:gpt-with-recv-vulnerability}
is similar to
\cref{fig:gpt-no-vulnerability}
, but
uses \mycode{recv} to receive user input.
The following shows the correct analysis result by \gpt 4.
\begin{quote}\it
    In the provided code snippet, the \mycode{recv}
    function is used as the taint source,
    and the second parameter (\mycode{a})
    is marked with the taint label for taint propagation.
    This setup introduces a potential vulnerability when the data received
    is copied from buffer \mycode{a} to buffer \mycode{b} using \mycode{strcpy},
    without checking the size of the data being copied.
     This scenario can lead to a buffer overflow vulnerability,
     specifically classified under CWE-120:
     Buffer Copy without Checking Size of Input (Classic Buffer Overflow).
\end{quote}

Based on the three analysis results provided by \gpt-4,
it is evident that \llms can understand and
leverage deep semantic information about the code
and the functions the code calls, to perform vulnerability checks.
Therefore, \llms have significant potential to automate
certain aspects of taint analysis, a topic that will be further elaborated in the remainder of this paper.

\begin{figure}[h]
    \centering
    \begin{minipage}{0.8\textwidth}
    \begin{lstlisting}[frame=leftline,xleftmargin=1em, numbers=left,numbersep=5pt,showstringspaces=false,basicstyle=\footnotesize\ttfamily,escapechar=|]
int |\textbf{main}|() {
    char a[50];
    char b[50];
    printf("Enter a string: ");
    recv(SOCKET, a, sizeof(a), 0); // The taint source changes
    strcpy(b, a); // A buffer overflow vulnerability here.
    return 0;
}
    \end{lstlisting}
    \end{minipage}
    \vspace*{-1em}
    \caption{A decompiled, vulnerable binary program that is the same as \cref{fig:gpt-no-vulnerability}
        except using \mycode{recv} to receive user input instead of \mycode{fgets} (comments are added to aid understanding).
        The reason for this program to have a buffer overflow vulnerability is that
        \mycode{recv} does not guarantee a null-terminator, a subtle behavioral difference from \mycode{fgets}.
        }
    \label{fig:gpt-with-recv-vulnerability}
    \end{figure}

\section{Overview of \proj}

\begin{lrbox}{\mybox}
	\begin{minipage}[*t]{\linewidth}
	\begin{lstlisting}[frame=none,numbers=none,basicstyle=\scriptsize\ttfamily]
	void (*@\textbf{FUN\_00100d40{}}@*)(void) {
	...
	iVar1 = socket(2,1,6);
	if (iVar1 != -1) {
		memset(local_88,0,16);
		local_88._0_2_ = 2;
		local_84 = inet_addr("127.0.0.1");
		local_88._2_2_ = htons(0x6987);
		iVar2 = connect(iVar1,(sockaddr *)local_88,0x10);
		if (iVar2 != -1) {
			sVar4 = recv(iVar1,(char *)((long)&local_78 +
			sVar3),99 - sVar3,0);
			...
		}
	}
	iVar1 = system((char *)&local_78);
	...
	}
	\end{lstlisting}
	\end{minipage}
	\end{lrbox}

We propose \proj, a novel approach that leverages the code understanding capabilities of
\llms to assist in taint analysis for identifying potential vulnerabilities in binaries.
This approach raises the key research question:
\textit{How can we practically and effectively prepare prompts to guide
\llms in performing vulnerability inspection?}
Effective prompts must provide \llms with sufficient code context
and clear analysis instructions. However, achieving this requires overcoming two main challenges.

\myparagraph{Challenge 1: Large Code Size and Complexity \vs Limited Token Context of \llms}
The size and complexity of binary code make it impractical to
directly feed an entire binary program to \llms for analysis.
\llms have limits on the number of tokens and the amount
of context they can process.
When these limits are exceeded,
\llms may lose track of previously analyzed information,
affecting subsequent code analysis. Even without these input limitations,
effectively inspecting vulnerabilities across the entire codebase
of a real-world software project remains a daunting challenge.

\myparagraph{Solution to Challenge 1}
Incorporating the principles of taint analysis,
we extract dangerous flows for \llms to analyze.
By slicing the program, we reduce the analysis complexity for \llms.
This method employs taint analysis principles to conduct
dependency analysis on dangerous data
related to the program's external input.
However, unlike traditional taint analysis,
we do not perform sanitization and vulnerability inspections;
instead, we focus solely on extracting dangerous flows.
\textit{A dangerous flow is defined as a slice of functions
involved in data dependencies between the external data input function
and potential destinations that could lead to vulnerabilities}
(see
\cref{subsec:call-chain-extraction} and
\cref{subsec:danger-flow-generation}).
Through this process, multiple dangerous flows can be extracted from the program,
forming a superset of potential vulnerabilities.
This approach partitions the analysis space for \llms,
reducing the overhead of complex taint analysis algorithms
and minimizing the risk of false negatives.
The extracted dangerous flows provide
\llms with relevant and contextually concise input,
enhancing the accuracy and efficiency of vulnerability analysis.

\myparagraph{Challenge 2: Obscurity of Vulnerability Discovery \vs Generality of Instructions for \llms}
Vulnerability discovery inherently involves exploring the unknown.
Instructing \llms to perform
similarity analysis based on known vulnerabilities
limits the ability of \llms to detect unknown vulnerabilities.
Additionally, it is impractical to provide
examples of all possible vulnerability patterns
in the prompt and use
few-shot learning for analysis~\cite{promptguide}.
The instructions for vulnerability checking
described in the prompt should remain general,
requiring no additional effort when the code or vulnerability types change.

\myparagraph{Solution to Challenge 2}
Inspired by the idea of manual code auditing
that usually splits a vulnerability analysis task into multiple smaller sub-tasks,
we instruct \llms to analyze the dangerous flow step by step in a similar manner.
When manually auditing  code,
we often use functions as units to gradually understand the entire program.
In addition, manual audits often determine whether vulnerabilities exist
in a general way by understanding the definition of functions and the processing flow of data,
rather than finding similarities through specific vulnerability examples.
Therefore, taking the functions in the dangerous flow as units,
we construct a comprehensive prompt sequence corresponding to the dangerous flow,
which specifies the data source to carry out data flow tracing and vulnerability inspection (see \cref{subsec:prompt-sequence-construction}).
This prompt sequence allows \llms
to complete the vulnerability analysis of
the dangerous flow in the form of a conversation.
Consequently, \llms
can explore and discover vulnerabilities more effectively,
even in scenarios with
different binaries or vulnerability types.

\subsection{Workflow of \proj}
\begin{figure*}[!htp]
	\centering
	\includegraphics[width=\linewidth]{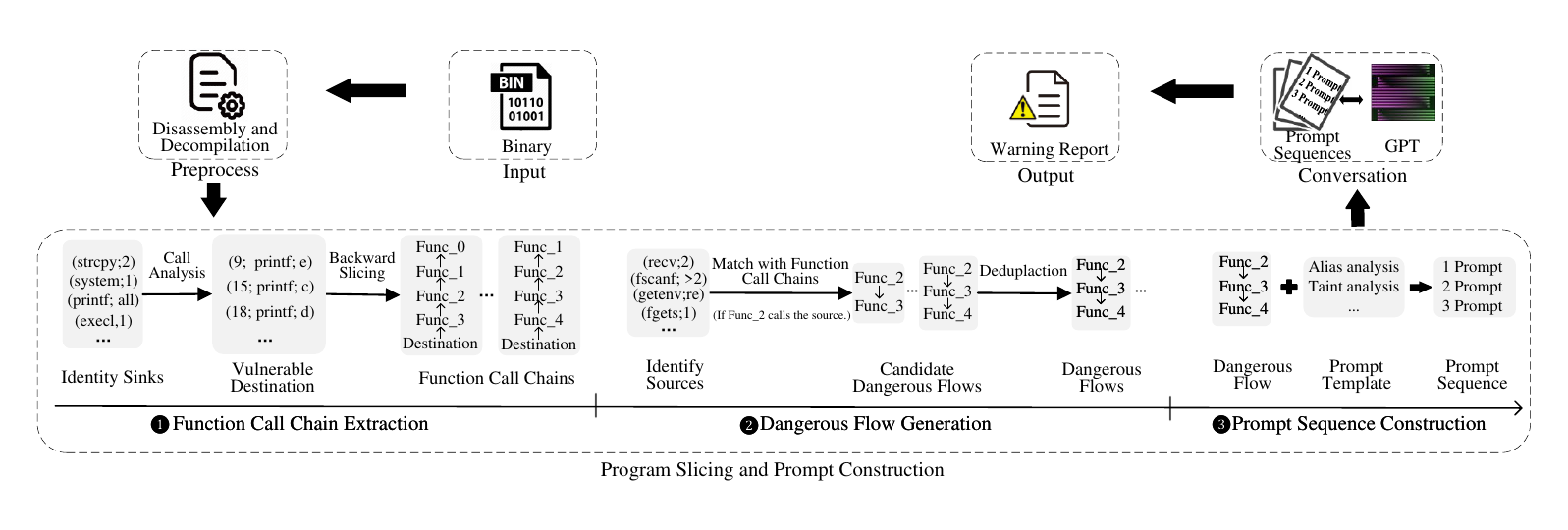}
	\caption{Workflow of \proj.}
	\label{fig-overview}
\end{figure*}

\proj utilizes the idea of \emph{code slicing} and
\emph{task splitting}
to build prompts to instruct \llms for vulnerability inspection.
The overall workflow of \proj is depicted in \cref{fig-overview},
primarily comprising three modules.
Firstly, we \textit{preprocess} the input binary
using disassembly and decompilation techniques.
This module provides
the recovered code structures for further analysis.
In the second module, we \textit{slice} the dangerous flows
of the decompiled program,
extracting relevant code snippets involving
data dependencies
between external inputs and vulnerable destinations.
For each dangerous flow,
we generate the corresponding \textit{prompt sequence}
that instructs the \llm.
Then, the third module uses the generated prompt sequence to engage in a \textit{conversation} with the \llm
to step-by-step guide
the dangerous flow analysis for discovering vulnerabilities.

The code slicing and prompt construction module serve as the core of \proj.
We further explain the three phases of this module.

\myparagraph{Phase 1:  Function Call Chain Extraction}
In this phase, we begin by identifying vulnerable destinations
(\ie, call sites of sinks)
within the binary.
To achieve this,
we utilize the \llm to analyze each external function to identify sinks and their related parameters,
which may potentially trigger vulnerabilities.
We consider the calls to these sinks as the vulnerable destinations.
Next,
we conduct backward data dependency analysis,
encompassing both intraprocedural and interprocedural aspects,
starting from each vulnerable destination.
This analysis helps us reconstruct the function call chains leading to these vulnerable destinations
(details in \cref{subsec:call-chain-extraction}).
Function call chains are defined as below.
\begin{definition}[Call Chain]
	A call chain \Algcallchain is a sequence of functions $[ f_1, f_2, \cdots, f_n  ]$, where
	$f_n$ contains a vulnerable destination \AlgVulnerdest and $\forall i\in[1, n): f_{i} \mbox{ calls } f_{i+1}$.
\end{definition}

\myparagraph{Phase 2: Dangerous Flow Generation}
The second phase focuses on identifying external input sources and matching them to the previously generated function call chains to generate dangerous flows.
To achieve this, we once again rely on the \llm to identify sources from external functions and determine their calls.
Then, we match the caller of the sources within the function call chains.
If the caller exists in a chain, we consider the chain as a candidate dangerous flow
whose starting point is the caller.
By applying a sequence-sensitive filtering process,
we eliminate repeated flows among candidate dangerous flows,
resulting in a set of distinct dangerous flows
(details in \cref{subsec:danger-flow-generation}).

\begin{definition}[Dangerous Flow]
	A dangerous flow \Algdangflow
	is a call chain $\Algcallchain = [ f_1, f_2, \cdots, f_n  ]$,
	where $f_1$ contains at least one call to a taint source,
    $f_i$ $(1 < i \le n)$ may contain calls to taint sources,
    $f_n$ contains a \AlgVulnerdest,
	and the external data obtained
	from any of the taint sources called on the chain
    can possibly flow to the \AlgVulnerdest.
\end{definition}

\myparagraph{Phase 3: Prompt Sequence Construction}
The third phase involves constructing prompt sequences
by combining the identified dangerous flows with specific analysis tasks.
For each dangerous flow,
we create the corresponding prompt sequence using functions as units.
To be applicable to various binaries,
 we design three prompt templates to complete the construction of prompts.
The first prompt template is used to initiate a conversation,
which combines with the first function in the flow to clarify the source of analysis,
and instructs the \llm to perform data aliasing, data dependency and taint analysis.
We then use a second template
combined with
each function in the dangerous flow to
recursively construct prompts
for data flow analysis until reach the end of the dangerous flow.
The last prompt instructs the \llm to initiate the final vulnerability analysis
(details in \cref{subsec:prompt-sequence-construction}).

\section{\proj Design}

\begin{algorithm}[ht]
	\DontPrintSemicolon
	\SetAlCapFnt{\fontsize{9pt}{15pt}}
	\small
	\caption{\fontsize{9pt}{15pt}Program Slicing and Prompt Construction.}
	\label{alg:extac}
	\KwIn{\AlgBinary: decompiled binary}
	\KwIn{\Algprompttemps: prompt templates}
	\KwOut{\AlgPromptSeqs: prompt sequences for vulnerability inspection}

	\tcp{Function Call Chain Generation}
		$\Algfuncext \gets \AlgExtractDynFuncNames(\AlgBinary)$\;
		$\Algextinpsinks \gets \emptyset$\;
		\ForEach{$\textit{f} \in \Algfuncext$}{
			\If{\llm thinks $f$ is a taint sink with the parameter(s) \AlgParam causing the vulnerability}{
				$\Algextinpsinks \gets \Algextinpsinks \cup \{ (f; \AlgParam) \} $ \label{line:main:sink:add}  \;
			}
		}

		vulnerable destinations \AlgVulnerdests $\gets \emptyset$\;
		\ForEach {$ (\Algextinpsink; \AlgParam) \in \Algextinpsinks $}{
			$\textit{callsites} \gets$ find all the call sites of \Algextinpsink \;
			\ForEach{$\AlgLoc \in \textit{callsites}$}{
                $\AlgArg \gets $ find the actual argument for the parameter \AlgParam\;
			    $\AlgVulnerdests \gets \AlgVulnerdests \cup \{ (\AlgLoc; \Algextinpsink; \AlgArg) \}  $\;
			}
		}

		call chains \Algcallchains $\gets \emptyset$   \label{line:main:cc:initial} \;
		\ForEach { \AlgVulnerdest $in$ \AlgVulnerdests }{
			$ \Algcallchain_1, \Algcallchain_2, ...  \gets$ \AlgBackwardslicing(\AlgBinary, \AlgVulnerdest) \label{line:main:cc:cal} \;
			$\Algcallchains \gets \Algcallchains \cup \{ \Algcallchain_1, \Algcallchain_2, ... \} \label{line:main:cc:uni}  $\;
		}
		\tcp{Dangerous Flow Generation}

		$\Algextinpsources \gets \emptyset$\;
		\ForEach{$\textit{f} \in \Algfuncext$}{
			\If{\llm thinks $f$ is a taint source with the parameter(s) \AlgParam storing external data}{
				$\Algextinpsources \gets \Algextinpsources \cup \{ (f; \AlgParam) \} \label{line:main:source:add}   $\;
			}
		}
		candidate dangerous flows \Algcandidatedangflows $\gets \emptyset$\;
		dangerous flows \Algdangflows $\gets \emptyset$\;
		\ForEach { \Algcallchain $\in$ \Algcallchains }{
			\If{ the \Algextinpsource of \Algextinpsources exists in \Algcallchain and has data dependency with the \AlgVulnerdest}{
				$\Algcandidatedangflow \gets $  change the starting function to the caller of  \Algextinpsource      \;
				$\Algcandidatedangflows \gets \Algcandidatedangflows \cup \Algcandidatedangflow   $\;
			}
		}
		$ \Algdangflows \gets$ \Algdeduplicate (\Algcandidatedangflows) \;

		\tcp{Prompt Sequence Construction}
		prompt sequences \AlgPromptSeqs $\gets \emptyset$\;
		\ForEach { \Algdangflow $\in$ \Algdangflows }{
			$\AlgPromptSeq \gets$ \Algconstruct(\Algprompttemps, \Algdangflow) \;
			$\AlgPromptSeqs \gets \AlgPromptSeqs \cup \AlgPromptSeq   $\;
		}
		return \AlgPromptSeqs \;
	\end{algorithm}

We introduce a novel technique for prompt sequence (\AlgPromptSeq) construction
based on dangerous flow (\Algdangflow) analysis,
which enables \llms to conduct effective vulnerability inspections in binaries.
\Algdangflows are obtained by performing backward data dependency slicing from
calls to \Algssfs
and matching \Algextinpsources.
The analysis of \Algdangflows focuses solely on data dependency analysis
and does not involve taint sanitation or vulnerability inspection.
Based on the extracted \Algdangflows,
combined with the task of program analysis,
the \llm inspects
\Algdangflows for potential vulnerabilities.
Combined with \cref{alg:extac},
we next explain the specific design of the program slicing and prompt construction process.

\subsection{Function Call Chain Extraction}\label{subsec:call-chain-extraction}

In the function call chain extraction phase,
we focus on extracting function call chains (\Algcallchains)
that are relevant to potential vulnerabilities in the binary.
This process involves two main steps:
identification of vulnerable destinations (\AlgVulnerdests)  and
backward slicing.

\subsubsection{Identification of Vulnerable Destinations }\label{ssfident}
The first step
is to find  the call sites of \Algssfs within the binary.
To inspect all potential vulnerabilities,
we need to locate all \Algssfs present in the binary.
Sinks often depend on external libraries linked to the binary,
as mentioned in \cref{sec:Motivation}.
To identify \Algssfs, we leverage the power of the \llm.
Specifically,
we extract externally linked functions
from
the decompiled binary (\AlgBinary) into a  list named \Algfuncext.
For dynamically linked libraries,
the names of external functions are preserved
in the function table of the binary,
so we can provide the function names to the LLM for analysis.
For statically linked libraries,
the body of the function is included in the binary.
We utilize the \llm for code summary analysis of the external library function to identify \Algssfs.
As shown in \cref{fig_ssfanalysis},
we take dynamic linking as an example to illustrate this process.
The form of static linking is just to replace the function name
with the function body.
The prompt first assigns the program analyst persona,
because the role setting performs better in statistical results~\cite{white2023prompt}.
Then, the prompt
clearly describes the purpose of the analysis,
\ie, determine whether the function can be used as a sink in taint analysis.
Moreover, we use the one-shot approach to give an example to standardize
the output of \llms.
The output of \llms is not necessarily completely accurate or stable.
We can remove the `without additional explanation' indication in the prompt
to use the description of the function in the response
to assist in the judgment of \Algssfs.
For any functions $f$ that is determined to be a sink,
we store the pair $(f; \AlgParam)$ of the function name
and parameter that needs to be checked,
into a set named \Algextinpsinks on \cref{line:main:sink:add} in \cref{alg:extac}
for subsequent use.
It is worth noting that the analysis results of
(\Algssf; \AlgParam) can be reused if another binary uses the same dynamic library.
This analysis is only necessary when
the binary uses a different dynamic library that has not been analyzed before.

\begin{figure}[!htp]
	\begin{tcolorbox}[colback=white, colframe=gray, title=Model Input, fonttitle=\bfseries]
		As a program analyst, is it possible to use a call \mycode{printf} as a
		sink when performing taint analysis? If so which
		parameters need to be checked for taint. Please answer yes
		or no without additional explanation. If yes, please indicate the
		corresponding parameters. For example, the \mycode{system} function
		can be used as a sink, and the first parameter needs to be
		checked as \mycode{(system; 1)}.
	\end{tcolorbox}

	\begin{tcolorbox}[colback=white, colframe=gray, title=Model Output, fonttitle=\bfseries]
		Yes, \mycode{printf} can be used as a sink. The parameters to be checked for taint are:

		Format string \mycode{(printf; 1)}

		Any additional parameters being formatted \mycode{(printf; 2, 3, ...)}
	\end{tcolorbox}
	\centering
	\caption{Use GPT-4.0 to identify sinks.}
	\label{fig_ssfanalysis}
\end{figure}

The output result is a set of pairs \mycode{(\Algssf; \Param)},
as shown on \cref{line:main:sink:add} in \cref{alg:extac}.
For example,
\mycode{(printf; 1), (printf; 2), $\cdots$} indicate
that every parameter of the \mycode{printf} function needs to be checked for
potential vulnerabilities.
After identifying the \Algssfs and the corresponding \AlgParam,
we traverse the binary to locate the call sites of each \Algssf,
forming the \AlgVulnerdest list of the triplet
\mycode{(\AlgLoc; \Algssf; \AlgArg)}.
For example,
the first parameter in the \mycode{printf} calls
in \cref{fig:integer_example} is a constant and does not need to be checked,
so \cref{fig:integer_example} contains three
VDs: \mycode{(9; printf; e)}, \mycode{(15; printf; c)}, and \mycode{(18; printf; d)}.

\begin{figure}[t]
  \centering
  \includegraphics[width=0.8\linewidth]{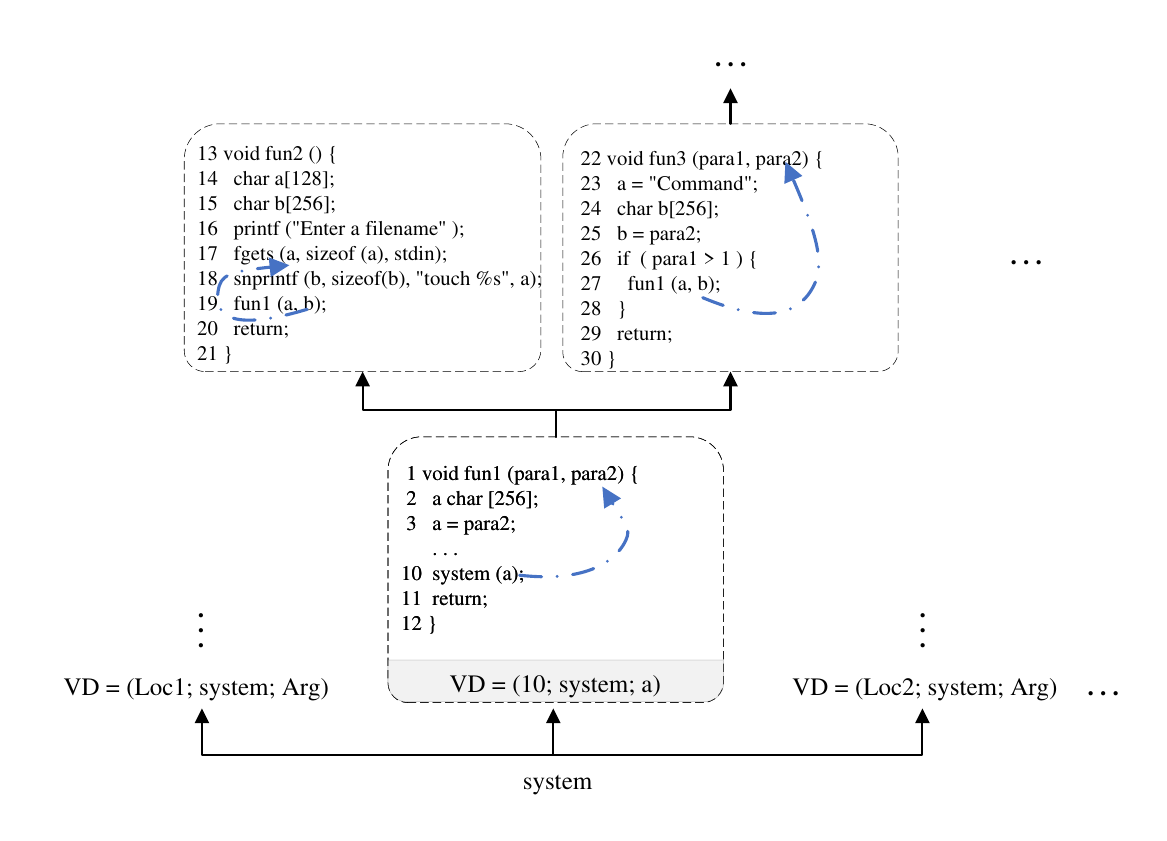}
\caption{
  An example call graph of a program, where each node is a function definition.
  Black arrows represent function call relationships.
  Blue arrows represent data dependency relationships.
  The figure directly shows the two call chains of
  $\textit{VD}= (\mycode{10}; \mycode{system}; \mycode{a})$,
  namely,
   $\{ [\mycode{fun2}, \mycode{fun1}], [\cdots, \mycode{fun3}, \mycode{fun1}]\}$.
}
\label{fig:callgraph}
\end{figure}

\subsubsection{Backward Slicing}

The second step, backward slicing,
involves obtaining the operations performed on the \AlgVulnerdest's
arguments before the call to the \Algssf.
This step relies on both intraprocedural and interprocedural data flow analyses.
We adopt a depth-first backward
approach to traverse the call graph
and analyze data flows
backward
from each
$\AlgVulnerdest=(\AlgLoc; \Algextinpsink; \AlgArg)$ to generate call chains
\Algcallchains, 
as shown on 
\ifInCopypaper
line 12–15 in Algorithm 1.
\else
\cref{line:main:cc:initial}--\ref{line:main:cc:uni}  in \cref{alg:extac}.
\fi

\proj  relies on
\ghidra~\cite{ghidra} to perform call graph analysis.
Specifically, we
obtain the address of each function\footnote{A function address
refers to the memory location where the code of the function begins in  memory.}
by traversing the functions identified by \ghidra.
Then, \mycode{call} and \mycode{jump}
instructions are found in each function,
and the memory addresses used in these instructions are
matched with the function addresses
to identify the relationship
between callers and callees
and further build the call graph.
Based on the call graph,
we firstly perform flow-insensitive and path-insensitive
intraprocedural data dependency analysis within
the function $f$ where \AlgVulnerdest is located.
For loops within $f$,
if duplicate paths are found,
the loop is exited.
In addition, we perform pointer alias analysis
with the help of variable description definitions and addressing methods~\cite{cheng2018dtaint}.
If direct data dependencies are detected between
the \AlgVulnerdest\unskip’s vulnerability trigger-related arguments
(\ie, \mycode{arg})
and the $f$’s parameters,
we extend our analysis to an interprocedural level.
The calling context is considered
by recording the function names
of different callers to form multiple call chains.
Therefore, we recursively analyze the dependencies
until there are no dependencies on caller parameters or no caller.
To manage the complexity of recursive calls and path explosion,
we limit the context call depth to 50.
Finally,
we add the call chains corresponding to different \AlgVulnerdests to the set \Algcallchains.

We use an example
to demonstrate the above analysis process.
\Algcallchains, 
as shown on 
\cref{fig:callgraph} 
shows
a partial call graph, where the library function \mycode{system} is the sink.
The $\AlgVulnerdest=(\mycode{10}; \mycode{system}; \mycode{a})$
is in $\mycode{fun1}$,
and we initialize its $\Algcallchain = [\mycode{fun1}]$.
There is a data dependency
between  \mycode{a}
and $\mycode{para2}$ of $\mycode{fun1}$.
So we continue to analyze the call sites of $\mycode{fun1}$ and
update \Algcallchains to
$\{ [\mycode{fun2}, \mycode{fun1}]$, $[\mycode{fun3}, \mycode{fun1}] \}$.
In $\mycode{fun3}$,
the
argument
corresponding to the parameter $\mycode{para2}$ of
$\mycode{fun1}$ is $b$,
and $b$ has a data dependency with $\mycode{para2}$ of $\mycode{fun3}$.
Therefore, we continue
to
analyze the callers of $\mycode{fun3}$
until there is no caller or no dependency on the parameters.

\subsection{Dangerous Flow Generation}\label{subsec:danger-flow-generation}
In the dangerous flow generation phase,
our goal is to identify \Algcallchains
in which external input
can reach a \AlgVulnerdest.
This process involves two main steps:
we first need to identify the external input source in the program
and then perform matching and deduplication in \Algcallchains
to determine \Algdangflows.

\subsubsection{Source Identification}
The first step is to identify \Algextinpsources in the binary.
Similar to the identification of \Algssfs in \cref{ssfident},
we also utilize the \llm for this task.
In this case, the analysis task is changed to find functions
that can directly receive external inputs.
We extract external link functions in the binary
and provide the function name or function body
so that \llm can perform auxiliary analysis
to determine whether the function
can be used as a taint source.
For any function $f$ that is determined to be a source,
we store the pair $(f; \AlgParam)$ of the function name
and parameter that needs to be checked
into a set named \Algextinpsources on \cref{line:main:source:add} in \cref{alg:extac}
for subsequent use.
For instance,
\cref{fig_sourceanalysis}
indicates that the \mycode{fscanf} function is used as the source,
and the second parameter stores the external input data.
Obviously, this response is not entirely accurate.
External input should be stored in the corresponding pointer
parameter after the format string.
Therefore, we need to make corrections based on slight expert experience.
We use \mycode{(fscanf; >2)}
to indicate that we need to analyze the parameters
after the second parameter of the \mycode{fscanf} function.
Like the identification of \Algssfs,
the source identification results can also be reused for
different binaries that uses the same dynamic library.

\begin{figure}[!htp]
\begin{tcolorbox}[colback=white, colframe=gray, title=Model Input, fonttitle=\bfseries]
  As a program analyst, is it possible to use a call to \mycode{fscanf} as a
  starting point (source) for taint analysis? If the function can
  be used as a taint source, which parameter in the call stores
  the external input data. Please answer yes or no without
  additional explanation. If yes, please indicate the corresponding
  parameters. For example, the recv function call can be used as a
  taint source, and the second parameter as a buffer stores the
  input data as \mycode{(recv; 2)}.
\end{tcolorbox}

\begin{tcolorbox}[colback=white, colframe=gray, title=Model Output, fonttitle=\bfseries]
  Yes, the \mycode{fscanf} function call can be used as a taint source,
  and the second parameter (format string) stores the input
  data \mycode{(fscanf; 2)}.
\end{tcolorbox}
  \centering
        \caption{Use GPT-4.0 to identify sources.}
         \label{fig_sourceanalysis}
  \end{figure}

\subsubsection{Matching and Deduplication}
With the identified set of \Algextinpsources $\{(\Algextinpsource; \AlgParam)\}$,
we proceed to match them in the extracted \Algcallchains
to identify \Algdangflows.
Specifically,
for each \Algcallchain,
we check whether the function $f$ of the \Algcallchain
call the identified \Algextinpsources.
If so, we further analyze whether
the argument corresponding
to
\AlgParam of the source
overlaps with the data flow of the \AlgVulnerdest.
If there is overlap,
the functions from $f$ to the end in the \Algcallchain
is considered a candidate dangerous flow (\Algcandidatedangflow).
As a \Algcallchain may generate multiple \Algcandidatedangflows due to multiple sources,
we only keep the longest \Algcandidatedangflow within the same \Algcallchain.
Additionally, different \Algcallchains may produce the same \Algcandidatedangflow or sub-chain.
To avoid redundancy and reduce unnecessary inspection requests,
we filter out \Algcandidatedangflows
that are sub-chains of other \Algcandidatedangflows.
Finally, we obtain a set of \Algdangflows
for the next step to construct prompt sequences.

We use $\Algcallchains= \{ [\mycode{fun2}, \mycode{fun1}], [\cdots, \mycode{fun3}, \mycode{fun1}] \}$
in \cref{fig:callgraph} to illustrate the above process.
In \mycode{fun2}, the taint source \mycode{fgets} is called
and the first argument \mycode{a} has a data dependency
with the  argument
\mycode{b} of \mycode{func1}
 (the \mycode{snprintf} in line 18 of \cref{fig:callgraph} stores \mycode{a} in \mycode{b}).
Therefore, $[\mycode{fun2}, \mycode{fun1}]$ is a \Algcandidatedangflow.
We assume that
$ \Algcallchain = [\cdots, \mycode{fun3}, \mycode{fun1}] $
forms $\{ [\mycode{fun5}, \mycode{fun4}, \mycode{fun3}, \mycode{fun1}], [\mycode{fun6}, \mycode{fun4}, \mycode{fun3}, \mycode{fun1}]  \} $
after the recursive analysis.
 If the taint source is called in \mycode{fun4} and \mycode{fun6},
 the corresponding \Algcandidatedangflows is $\{ [\mycode{fun4}, \mycode{fun3}, \mycode{fun1}], [\mycode{fun6}, \mycode{fun4}, \mycode{fun3}, \mycode{fun1}]  \} $.
 But $[\mycode{fun4}, \mycode{fun3}, \mycode{fun1}]$
 is a subchain of $ [\mycode{fun6}, \mycode{fun4}, \mycode{fun3}, \mycode{fun1}] $,
 so the former is discarded.
 The final \Algdangflows are $\{ [\mycode{fun2}, \mycode{fun1}], [\mycode{fun6}, \mycode{fun4}, \\ \mycode{fun2}, \mycode{fun1}]  \} $.

\begin{figure}[h]
\begin{subfigure}[b]{\textwidth}
\centering
\begin{tcolorbox}[colback=white, colframe=gray, title=Template for Start Prompt, fonttitle=\bfseries]
As a program analyst,
I give you snippets of C code generated by decompilation, using \mycode{<function>}
as the taint source, and the \mycode{<parameter>} parameter marked as the taint label to extract the taint data flow. Pay attention to the data
alias and tainted data operations. Output in the form of data flows.

\mycode{<Code to be analyzed>}

\end{tcolorbox}
\caption{Template for the start prompt.}
\label{tem:subfig1}
\end{subfigure}

\begin{subfigure}[b]{\textwidth}
\begin{tcolorbox}[colback=white, colframe=gray, title=Template for Middle Prompt, fonttitle=\bfseries]

Continue to analyze function according
to the above taint analysis results.
Pay attention to the
data alias and tainted data operations.
(Note the new taint source \mycode{<function>} and the \mycode{<parameter>} parameter marked as the taint label.)

\mycode{<Code to be analyzed>}

\end{tcolorbox}

\caption{Template for the middle prompt.}
\label{tem:subfig2}
\end{subfigure}

\begin{subfigure}[b]{\textwidth}

\begin{tcolorbox}[colback=white, colframe=gray, title=Template for End Prompt, fonttitle=\bfseries]
Based on the above taint analysis results, analyze whether the code has vulnerabilities. If there is a
vulnerability, please explain what kind of vulnerability according to CWE.

\end{tcolorbox}

\caption{Template for the end prompt.}
\label{tem:subfig3}
\end{subfigure}

\centering
\caption{The prompt templates for constructing the PS that instruct the LLM to inspect vulnerabilities.}
\label{fig-templete}

\end{figure}

\subsection{Prompt Sequence Construction}\label{subsec:prompt-sequence-construction}
In the prompt sequence construction phase,
our objective is to create prompt sequences (\AlgPromptSeqs)
that instructs the LLM to perform vulnerability
inspection based on identified
\Algdangflows.
To ensure that the \AlgPromptSeq effectively instructs the \llm
and remains valid and effective for different, diverse programs,
we propose prompt templates (\Algprompttemps)
that are applicable to all \Algdangflows and can be instantiated to fit each \Algdangflow.
We follow a rigorous iteration and optimization process
that ranged from initial prompt design to testing and feedback analysis.
The final PTs include the following key elements for effective analysis:
\begin{itemize}[leftmargin=10pt,topsep=0pt]
    \item Clearly state the analysis tasks.
    Precise instructions for vulnerability analysis are essential to prevent irrelevant content.
    \item Enrich the context of tasks.
    Additional instructions related to vulnerability analysis,
    such as alias analysis and focusing on tainted data flows, enhance accuracy.
    \item Persona assignment.
    Assigning appropriate roles improves the professionalism of \llms
    in responding to descriptions.
    \item Splitting of analysis tasks or data.
    For large data to be analyzed,
    splitting or preprocessing the data helps improve accuracy.
    \item The specified output format.
    The determined output format facilitates analysis for downstream tasks.
  \end{itemize}
  According to these elements,
  we design three templates shown in the
\cref{fig-templete}.
The PTs are inspired by the general approach of manual code auditing.
Specifically, data dependency analysis starts from the function containing the external data source
and progresses through the function call chain until the last function in the \Algdangflow.
If a \Algdangflow contains only one function, then
the \AlgPromptSeq for it includes a start prompt and an end prompt.
For a \Algdangflow consisting of more than one function, the middle template is used recursively to complete the PS construction.
This structured approach ensures that LATTE performs vulnerability analysis effectively and consistently across different scenarios.
The detailed explanation of \Algprompttemps design is as follows.

\myparagraph{Start Template (\cref{tem:subfig1})}
Initializes the analysis by specifying the task and context clearly.
The start prompt assigns the persona of the \llm
and explains the analysis background and tasks.
Specifically, we instruct the \llm to conduct the following conversion
in the role of program analyst.
Then, we clarify the taint source of program analysis.
For example, for the $\Algdangflow = [\mycode{fun2}, \mycode{fun1}]$ shown in \cref{fig:callgraph},
the pair  (\mycode{<function>}, \mycode{<parameter>})
in \cref{tem:subfig1} is replaced with (\mycode{fgets}, \mycode{first}).
At the same time,
the body of first function \mycode{fun2}
is appended at the end of the prompt for analysis.
In the description of the analysis task,
we mainly focus on the propagation of tainted data,
so we emphasize the analysis of data dependencies.
To conduct more accurate analysis,
we also emphasize the focus on data aliases and tainted data operations.

\myparagraph{Middle Template (\cref{tem:subfig2})}
Recursively used to handle intermediate steps in the \Algdangflow,
enriching the context and focusing on relevant data flows.
Specifically,
the middle prompt are reused to describe functions in a \Algdangflow other than the first function.
Thanks to the short-term memory capability of the \llm itself,
we can complete data dependency analysis between functions.
Therefore, the middle prompt needs to indicate
that the taint analysis is continued
based on the analysis results of the previous function.
Similarly, we put the function body to be analyzed at the end of the prompt.
For example,
for the $\Algdangflow = [\mycode{fun2}, \mycode{fun1}]$ shown in \cref{fig:callgraph},
the body of function \mycode{fun1} is appended at the end of the prompt for analysis.
Moreover, if there is a \Algextinpsource in the current function,
we fill in the source name and the corresponding parameter
at \mycode{<function>} and \mycode{<parameter>}
as in the start template.
Otherwise, taint propagation analysis is performed
based on the analysis results
of the previous function
and the prompt
does not include the content in parentheses (\ie,
``(Note the new taint source ... as the taint label.'' in \cref{tem:subfig2}).

\myparagraph{End Template (\cref{tem:subfig3})}
Concludes the analysis, ensuring that all necessary information is captured and formatted correctly.
Specifically
the end template is used after all functions in have been analyzed.
The main task of the end template is to perform vulnerability inspection
based on the analysis results of all \Algdangflow functions.
At the same time, to facilitate vulnerability identification and analysis,
we also let the end template  classify vulnerabilities based on CWE.
It is worth mentioning that this not only provides a more detailed vulnerability analysis description,
but also provides patching suggestions based on the code context.

Assuming that the DF only contains the \mycode{foo} function in \cref{fig:integer_example},
the running example
of \proj instructing the \llm to perform vulnerability analysis is shown in \cref{fig-cover}.
First, the output of \llm clearly describes the taint propagation process.
The model then gives the location and reason why the vulnerability was triggered
in the output of the vulnerability analysis.
Therefore, we can not only see the vulnerability check results,
but also further verify the vulnerability or patch the vulnerability
based on the analysis process in the response of \llms.

\begin{figure}[!htp]
	\footnotesize
\begin{tcolorbox}[colback=white, colframe=gray, title=Model Input, fonttitle=\bfseries]
  As a program analyst, I give you snippets of C code generated by decompilation, using fscanf as the taint source, and third parameter
  marked as the taint label to extract the taint data flow. Pay attention to the data alias and tainted data operations. Output in the form of
  data flows.

void foo(void) \{

\quad char a, b, c, d;

\quad …

\end{tcolorbox}

\begin{tcolorbox}[colback=white, colframe=gray, title=Model Output, fonttitle=\bfseries]
  To analyze the data flow in the provided C code, I'll focus on the taint source, which is the fscanf function using the third parameter ($\&$a) as the
  taint label. I'll also take into account the data aliasing and tainted data operations. Here's a step-by-step analysis of the data flow:

  1: fscanf(stdin, "\%d", $\&$a); - a is tainted by the input from stdin.

  2: b = a; - b gets tainted with the same value as a.

  3: c = 0; - c is not tainted.

  4: a = 1; - a is no longer tainted, overwritten with a constant value.

  5: e = b + 1; - e is tainted by the value of b, which is tainted.

  6: printf("\%d",(ulong)e); - e is printed, but it's cast to ulong.

  7: if (b == 127) - Condition checks tainted value of b.

  If true:

  ...
\end{tcolorbox}

\begin{tcolorbox}[colback=white, colframe=gray, title=Model Input, fonttitle=\bfseries]
  Based on the above taint analysis results, analyze whether the code has CWE-190 integer overflow vulnerabilities.
\end{tcolorbox}

\begin{tcolorbox}[colback=white, colframe=gray, title=Model Output, fonttitle=\bfseries]
  The code snippet you provided does indeed contain a potential integer overflow vulnerability, specifically related to CWE-190 (Integer
  Overflow or Wraparound).

  Let's examine the vulnerable part of the code: d = c + 1;

  Here, c is potentially tainted by the input value of b. If b is close to the maximum value that a char can hold (which is typically 127 for signed
  char), then adding 1 to it (c + 1) can result in an overflow condition, leading to undefined behavior.

  Although c is assigned the value of b + 1 in an earlier part of the code (c = b + 1;), the subsequent addition of 1 to c (d = c + 1;) is where the
  overflow may occur.

  …
\end{tcolorbox}

  \centering
  \caption{
    Based on the $\textit{foo}$ function in \cref{fig:integer_example},
    the process of using GPT-4.0 to analyze and check bugs is shown.
    The final model output correctly identifies that the $d = c + 1$
    operation of the $\textit{foo}$  has a potential integer overflow vulnerability.
  }
  \label{fig-cover}

  \end{figure}

\section{Evaluation}\label{sec:evaluation}

The evaluations of \proj are designed to address the following five research questions:
\begin{description}[leftmargin=*,topsep=0pt]
  \item [RQ1:] How well does \proj
    inspect vulnerabilities in binaries?
  Explain from the following three perspectives:
  \begin{enumerate}[leftmargin=20pt, topsep=0pt]
   \item How effective is \proj in identifying security vulnerabilities?
   \item How stable is  \proj in vulnerability inspection result?
   \item How efficient is \proj in terms of processing time and monetary cost?
  \end{enumerate}

    \item [RQ2:] How effective is the identification of security sensitive functions (\Algssfs)
  and external input sources (\Algextinpsources)?
  \item [RQ3:] How effective is the extraction of dangerous flows (\Algdangflows)?

  \item[RQ4:] What is the performance impact of prompt sequence (\AlgPromptSeq) on \proj?

  \item [RQ5:] Does \proj scale to real-world binary vulnerability analysis?

\end{description}

\subsection{Evaluation Setup}

\myparagraph{Implementation}
The implementation of \proj is based on the reverse engineering framework Ghidra~\cite{ghidra,Ghost}
and the large language model GPT-4.0~\cite{GPT4}.
Below, we outline the overall implementation process.
The first step involves loading the target binary into Ghidra, 
where the custom Ghidra plugin is utilized to automate extraction tasks. 
Through the plugin, 
LATTE retrieves critical information from the binary, 
such as decompiled code, the call graph, 
control flow graph (CFG), import table, 
and additional metadata necessary for in-depth analysis.
Once the binary data is collected, 
LATTE conducts detailed program analysis to extract DFs 
that are relevant to vulnerability triggering. 
Using predefined prompt templates, 
LATTE constructs PSs that correspond to each DF. 
Each PS is specifically designed to interact with GPT-4.0, 
setting up a session in which the \llm 
conducts a vulnerability assessment of the associated DF.
As GPT-4.0 processes each PS, 
it evaluates the corresponding DF for potential vulnerability issues. 
Once the analysis across all DFs within a binary is completed, 
LATTE saves the results generated by GPT-4.0. 
These results, stored as individual files, 
are organized for further verification and validation, 
supporting a comprehensive review of any identified vulnerabilities.

\myparagraph{Benchmarks}
To thoroughly and fairly evaluate the capabilities of \proj,
we used the Juliet Test Suite~\cite{juliet}
and \karonte data set~\cite{redini2020karonte}.
 We first selected the benchmarks included in the standard data set \juliet Test Suite (v1.3)
according to \cref{tab:cweandtainter} to evaluate RQ1-RQ4
(\juliet has no benchmarks for CWE-120 and we used the \karonte dataset for CWE-120).
To ensure that our evaluation mirrors real-world scenarios, we removed debug information and symbol tables from the compiled benchmark programs.
Furthermore, since \proj aims to detect vulnerabilities caused by external inputs,
test subjects with internal data as a source have been removed.
The number of test subjects used for evaluation
is listed in \cref{tab:juliet-eval}.
Second, we performed RQ5 evaluation
on real firmware data set provided by Karonte
to illustrate \proj's ability to find vulnerabilities in the real world.
Since the source code of firmware in Karonte
are not publicly accessible,
 and thus not in the training sets of \llms.
 The evaluation on Karonte helps to investigate whether
 \proj suffers from the data leakage problem~\cite{leakage}.

\myparagraph{Baselines}
We selected Emtaint~\cite{cheng2023emtaint} and Arbiter~\cite{vadayath2022arbiter}
for comparison on the Juliet Test Suite.
These techniques represent the state of the art as shown
in
\cref{tab:cweandtainter}.
For the Karonte dataset,
we used Karonte~\cite{redini2020karonte} as another baseline for comparison,
because Karonte is designed specifically for analyzing fimrware.
We did not include CodeQL~\cite{codeql}, QueryX~\cite{QueryX}
or Joern~\cite{joern}
as baselines due to their reliance on
customized queries for bug detection~\cite{QueryX}.

\myparagraph{Metrics}
In the evaluation based on the Juliet Test Suite,
we measured the effectiveness of \proj using six indicators:
true positive (\tp), false negative (\fnn), true negative (\tn),
false positive (\fp),
accuracy ($ \frac{\textit{\tp} + \textit{\tn}}{\textit{\tp} + \textit{\tn} + \textit{\fp} + \textit{\fnn}} $),
and \fscore score ($ \frac{2 * \textit{precision} * \textit{recall}}{\textit{precision} + \textit{recall}}$,
where $\textit{precision} = \frac{\textit{\tp}}{\textit{\tp}+\textit{\fp}}$, $\textit{recall} = \frac{\textit{\tp}}{\textit{\tp} + \textit{\fnn}}$).
For the evaluation on the real-world firmware dataset provided by \karonte,
since there are no labels,
we mainly measured the \tp and \fp in the analysis alerts.

\myparagraph{Running Environment}
All experiments were performed on a Linux workstation
with an Intel Core i7-8750H CPU and 64G RAM.

\subsection{Vulnerability Inspection (RQ1)}

\begin{table*}[t]
  \caption{Evaluation results of vulnerability inspection based on Juliet. \ding{53} indicates that the vulnerability type is not supported.
  }
  \label{tab:juliet-eval}
  \begin{threeparttable}
  \resizebox{\textwidth}{!}{%
  \begin{tabular}{@{}lllllllllllll@{}}
      \hline
      \multirow{2}{*}{} & \multicolumn{3}{c}{CWE-78 (960/960)$^+$ }            & \multicolumn{3}{c}{CWE-134 (1200/1200)$^+$}            & \multicolumn{3}{c}{CWE-190 (2860/2860)$^+$}          & \multicolumn{3}{c}{CWE-606 (240/240)$^+$}  \\ \cline{2-13}
                        & Emtaint   & Arbiter & \proj          & Emtaint        & Arbiter & \proj        & Emtaint & Arbiter$^*$ & \proj             & Emtaint & Arbiter & \proj    \\ \hline
      TP                & 820       &    408  & \cellcolor[HTML]{C0C0C0}892    & 1115  &     1166    & \cellcolor[HTML]{C0C0C0}1151       & \ding{53}       &    2091     & \cellcolor[HTML]{C0C0C0}1773   &\ding{53} &\ding{53}  & \cellcolor[HTML]{C0C0C0}210\\
      FN                & 140       &   552     & \cellcolor[HTML]{C0C0C0}68    & 85     &      34   & \cellcolor[HTML]{C0C0C0}49              & \ding{53}       &    769     & \cellcolor[HTML]{C0C0C0}1087        &\ding{53} &\ding{53}  &   \cellcolor[HTML]{C0C0C0}30 \\
      TN                & 960       &    430    & \cellcolor[HTML]{C0C0C0}960   & 1132   &     30   &      \cellcolor[HTML]{C0C0C0}1102       &   \ding{53}      &   444      &   \cellcolor[HTML]{C0C0C0}1779          &\ding{53} & \ding{53} &   \cellcolor[HTML]{C0C0C0}142  \\
      FP                & 0         &    530    & \cellcolor[HTML]{C0C0C0}0     & 68    &     1170    &     \cellcolor[HTML]{C0C0C0}98        &    \ding{53}     &   2416      &   \cellcolor[HTML]{C0C0C0}1081       &\ding{53} &\ding{53}  &   \cellcolor[HTML]{C0C0C0}98    \\
      Accuracy         & 92.7\%     &    43.64\% & \cellcolor[HTML]{C0C0C0}96.46\%   & 93.63\%       &     49.83\%   & \cellcolor[HTML]{C0C0C0}93.88\%       &   \ding{53}   &   44.32\%   &   \cellcolor[HTML]{C0C0C0}62.1\%    &\ding{53} &\ding{53}  &   \cellcolor[HTML]{C0C0C0}73.33\%  \\
      Precision         & 100\%     &    43.5\% & \cellcolor[HTML]{C0C0C0}100\%   & 94.25\%       &     49.91\%   & \cellcolor[HTML]{C0C0C0}92.15\%       &   \ding{53}   &   46.39\%   &   \cellcolor[HTML]{C0C0C0}61.12\%    &\ding{53} &\ding{53}  &   \cellcolor[HTML]{C0C0C0}65.26\%  \\
      Recall        & 85.42\%     &    42.5\% & \cellcolor[HTML]{C0C0C0}92.92\%   & 92.92\%       &     97.17\%   & \cellcolor[HTML]{C0C0C0}95.92\%       &   \ding{53}   &   74.68\%   &   \cellcolor[HTML]{C0C0C0}68.19\%    &\ding{53} &\ding{53}  &   \cellcolor[HTML]{C0C0C0}87.5\%  \\
      F1 Score          & 92.14\%    &    43.25\% & \cellcolor[HTML]{C0C0C0}96.33\%    & 93.58\%      &     65.95\%   &  \cellcolor[HTML]{C0C0C0}93.99\%       &   \ding{53}   &   56.76\%   &  \cellcolor[HTML]{C0C0C0}62.05\%       &\ding{53} & \ding{53} & \cellcolor[HTML]{C0C0C0}74.24\%    \\
      Avg. Time (s)      & 5.2    &    122 & \cellcolor[HTML]{C0C0C0}13.5    & 5.6      &     73.6  &  \cellcolor[HTML]{C0C0C0}14.2      &   \ding{53}   &   7.8   &  \cellcolor[HTML]{C0C0C0}15.1       &\ding{53} & \ding{53} & \cellcolor[HTML]{C0C0C0}13.9    \\ \hline
    \end{tabular}%
  }
  \begin{tablenotes}
      \footnotesize
      \item[+] \textit{Number of test subjects containing vulnerabilities} / \textit{Number of test subjects not containing vulnerabilities}
      \item[*] Arbiter relies on symbol table, such as \mycode{printIntLine}, \mycode{printHexCharLine}, and \mycode{printLongLongLine}, to assist \\ in overflow judgment. If \proj is also analyzed based on the binary
      with the symbol table, TP is 2576, FN is 284,\\ TN is 1939, FP is 921, Precision is 73.66\%, Accuracy is 78.93\%, Recall is 92\%, F1 is 81.04\%.
    \end{tablenotes}
  \end{threeparttable}
  \end{table*}

\subsubsection{Effectiveness}\label{subsec:effect}
\proj performs vulnerability inspection
on stripped binaries compiled from Juliet.
The name of the test subject
describes the \cwe vulnerability types it contains.
However, there are overlaps between \cwe
vulnerability types (\cref{subsec:vulnerability-patterm}).
To facilitate the verification of the results,
we only determine whether the \cwe vulnerability type
specified by the test subject exists.
In evaluating \proj,
we performed five rounds of testing
to ensure robustness and accuracy.
As shown in \cref{tab:juliet-eval},
the inspection accuracy of the four types of vulnerabilities
CWE-78, CWE-134, CWE-190 and CWE-606 are 96.46\%, 93.88\%, 62.1\% and 73.33\% respectively,
and the average accuracy is 77\%.

\proj outperforms both \emtaint and \arbiter,
in terms of accuracy and \fscore score.
Note that in the case of CWE-190 (Integer Overflow or Wraparound),
\arbiter can detect
318
more \tps than \proj.
However, this is attributed to the fact that \arbiter
relies on the symbol table for inspection,
and if this information is provided to \proj,
\proj can outperform \arbiter by detecting 485
more \tps.

\myparagraph{Analysis of \fnns and \fps}
We further analyzed the reasons behind \fnns and \fps
in \proj vulnerability checks.
The first reason is that the dangerous flows of test subjects
are not always correctly extracted.
This may occur due to program analysis problems,
such as indirect calls and pointer aliases,
which can lead to false or failed extraction of dangerous flows (details in \cref{sec:dangerous}).
The second reason is related to the creativity of \gptfour
which can lead to different responses to the same prompt sequence,
resulting in unstable analysis results.
However, the vulnerability inspection task itself has the property of exploration
rather than simple rule matching.
Therefore, the instability brought about by the creativity of \gptfour
is not necessarily bad for vulnerability inspection.
We further analyzed the stability of the \proj result (details in \cref{subsec:stability}).

\begin{table}[t]
	\centering
	\caption{Comparison of vulnerability checking accuracy of \proj based on \gptfour and \gptthree.}
	\label{tab:modelcompare}
		\begin{tabular}{@{}lll@{}}
			\hline
			& \gptfour & \gptthree \\ \hline
			CWE-78   & 96.46\%            & 78.2\%             \\
			CWE-134  & 93.88\%            & 75.9\%             \\
			CWE-190  & 62.1\%             & 54.5\%             \\
			CWE-606  & 73.33\%            & 61.2\%             \\ \hline
			Average  & 81.44\%            & 67.45\%             \\ \hline
		\end{tabular}%
\end{table}

\myparagraph{\gptfour \vs \gptthree}
\proj based on \gptfour
can perform effective taint-style vulnerability inspection.
We further used \gptthree to conduct an ablation experiment on the model.
\cref{tab:modelcompare} shows
that \proj based on \gptfour performs better than \gptthree
on each vulnerability type.
The average accuracy of \proj
based on \gptfour is better than that of \gptthree 13.99\%.
This illustrates that the ability of the model itself directly affects
the analysis effect of \proj.
With the development of \llms,
the effectiveness of vulnerability checks will also increase.
The design of \proj allows easy switching between different \llms.

\subsubsection{Stability}\label{subsec:stability}
As an artificial intelligence generation model,
\gptfour exhibits a certain degree of randomness in the generated response.
The temperature parameter plays a significant role in controlling the creativity
and diversity of the content produced by \gptfour.
The temperature value is a floating-point number between \mycode{0} and \mycode{1},
where higher values result in more diverse and random outputs,
while lower values produce more conservative and deterministic outputs.
We conducted two sets of experiments
to evaluate the stability of \proj analysis results
and the effect of temperature on the results.

\begin{figure}[t]
  \centering
  \includegraphics[width=0.8\linewidth]{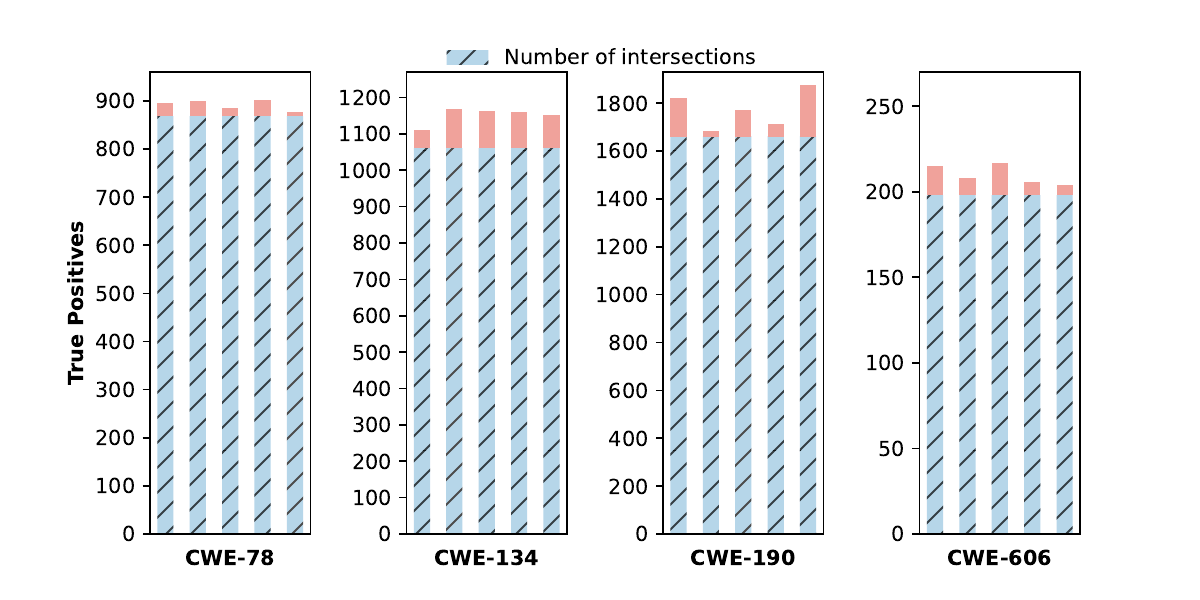}
  \caption{The \tp intersection among the five test rounds indicates the stability of \proj. The intersection numbers of the four vulnerability types are 868, 1061, 1660, 198.}
  \label{fig-stability}
\end{figure}

\myparagraph{Analysis Stability at Temperature 0.5}
To strike a balance between creativity and stability,
we used a median temperature value of \mycode{0.5} in the five rounds of testing in \cref{subsec:effect}.
We measured the stability by evaluating the intersection of \tps among the five rounds.
\cref{fig-stability} shows small fluctuations
in the number of \tps in different rounds of the same CWE type.
From left to right in the \cref{fig-stability},
the recall rates of five-round intersection
are 90.42\%, 88.42\%, 58.04\%, and 82.5\%, respectively,
(the average recall rate is 79.85\%).
This indicates that the stability of \proj is acceptable.
However, we noticed relatively large fluctuations in the CWE-190.
The reason for the fluctuation of CWE-190
is consistent with the problem mentioned above (\cref{subsec:effect}), which is
caused by incorrect data type judgments without the assistance of function names.

\begin{figure}[!htp]
	\centering
	\includegraphics[width=0.7\linewidth]{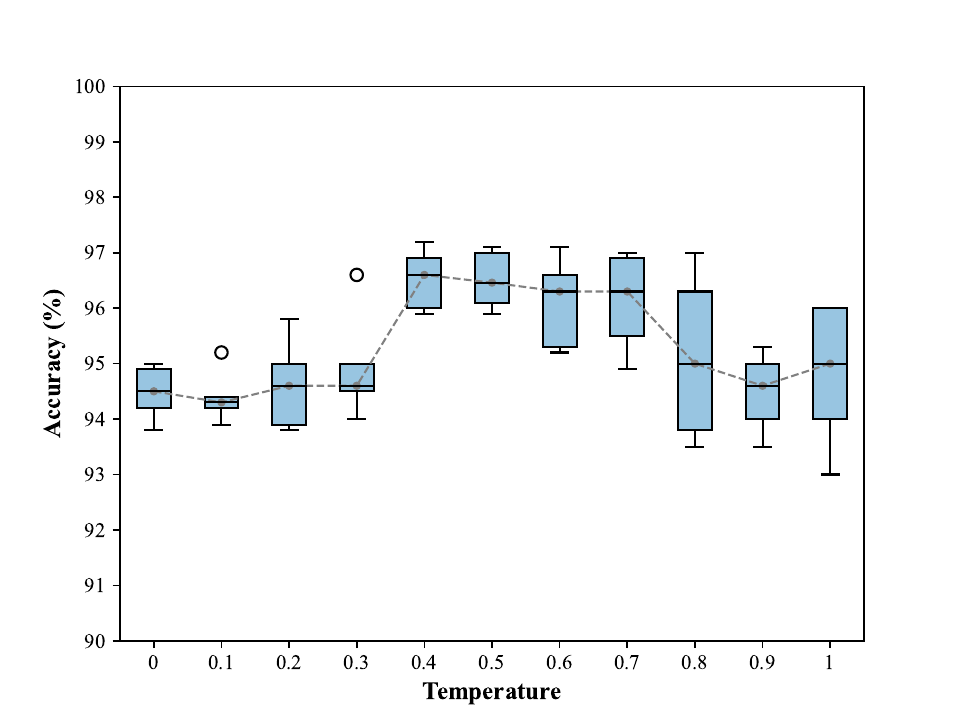}
	\caption{Accuracy of CWE-78 analysis for \proj at different temperatures.}
	\label{fig-temperature}
\end{figure}

\myparagraph{Impact of Temperature on Analytical Results}
To explore the impact of temperature on \proj,
we conducted 11 experiments with different temperature values.
Each experiment was performed five rounds to ensure reliability.
\cref{fig-temperature} shows the accuracy results of CWE-78,
where the trend generally follows a normal distribution,
including the median, minimum, and maximum values.
Moreover, we observe that when the temperature is less than \mycode{0.4},
the inspection results have a smaller distribution range.
Conversely, when the temperature is greater than \mycode{0.6},
the inspection results are more scattered.
The temperature of \mycode{0.4-0.6} can achieve a balance between stability and inspection effect.
Similar trends were observed for the analysis of CWE-134, CWE-190, and CWE-606.

\subsubsection{Efficiency}
We analyzed the time overhead and the monetary cost of
\proj when inspecting for vulnerabilities.
This process refers to the overhead of binary loading,
dangerous flow extraction, and conversation with GPT-4.0 for vulnerability inspecting.
Source and sink results can be reused between different binaries so they are preconfigured
(evaluated in
  \cref{source-sink-iden}).

The average time taken to analyze different test subjects is summarized in \cref{tab:juliet-eval}.
The results indicate that \proj can complete the analysis of a test subject within 20 seconds.
Preprocessing the test subject and extracting the dangerous flows takes about two seconds,
which is a relatively efficient process.
The majority of the analysis time is spent on the conversation with \gptfour,
as this step involves instructing the model and waiting for its response.
The time taken for this step can be influenced by network quality and the response speed of GPT-4.0.
Comparing \proj with \emtaint and \arbiter,
we find that the time overhead of \proj is slightly greater than that of \emtaint,
with a difference of about 8 seconds.
This difference is considered acceptable
given the automation that comes with using \gptfour.

Regarding the monetary cost, the evaluated cost of \proj analyzing a dangerous flow is
\$ 0.35,
which is considered acceptable compared to the expenditure of hiring a program analysis expert for manual inspection.
Moreover, with the development of \llm technology,
the cost of using such models is expected to decrease over time,
while the performance is likely to improve,
making it an even more attractive option for vulnerability analysis.

  \finding{The average accuracy of \proj vulnerability inspection is 77\%.
  Compared to the state-of-the-art techniques Emtaint and Arbiter, \proj outperforms them in different vulnerability types.
  Moreover, the average recall of the \tp intersection across different test rounds is 79.85\%,
  which shows that the inspection results of \proj are stable.
  }

\subsection{Sink and Source Identification (RQ2)}\label{source-sink-iden}

\proj reasonably identified \Algssfs and \Algextinpsources as necessary for dangerous flow extraction.
As shown in \cref{tab:my-srcdst},
we verified the results of \Algssf and \Algextinpsource identified by \proj.
For each test subject,
the vulnerability is triggered by a specific pair of \Algssf and \Algextinpsource,
such as \mycode{system} and \mycode{recv}.
The expected pair of the test subject
 is indicated by its name. 
 \proj accurately covers 
 both the \Algssf and \Algextinpsource
 of the target scenario in the test subject with 100\%.
The time overhead for identifying sinks and sources
is proportional to the number of library functions.
For the test subject in Juliet, 
48 queries were required, 
with each query taking approximately \mycode{1} seconds.
Furthermore, these results can be cached for all test subjects 
from Juliet without the need for further queries.

There are other functions in a test subject 
that can be used as \Algssfs and \Algextinpsources,
but they do not cause a vulnerability in the current test subject.
The identified functions can be used as \Algssfs and \Algextinpsources
with average precision up to 82.1\%.
FPs are mainly attributed to the instability of GPT-4.0.
For example, sometimes \mycode{recv} is mistakenly treated as an \Algssf.
This highlights the challenge of managing the creativity and diversity of the \llm,
which can sometimes result in unintended outcomes.
To alleviate this situation,
we can analyze function semantics to filter FPs and utilize TPs
to form \Algssf and \Algextinpsource collections to assist analysis.
When evaluating real programs in
\cref{realsec},
we use the correct \Algssf and \Algextinpsource
results identified from the test subjects to analysis.

Compared with the state-of-the-art binary taint analysis techniques,
such as \emtaint and \arbiter,
\proj improves the automation of taint analysis.
Previous techniques primarily relied on manual analysis of function semantics 
to identify
\Algssfs and \Algextinpsources,
making it difficult to compare identification results. 
However, the ability of \llms 
to understand code and generate function summaries
can be used to assist manual analysis and reduce manual effort.
\proj simplifies the analysis process
and improve the efficiency of
existing techniques
by providing analysts
with more program information.

\begin{table}[!htb]
  \caption{Sink and source identification results}
  \label{tab:my-srcdst}
  \small
  \centering
  \begin{threeparttable}
  \begin{tabular}{@{}lllllll@{}}
  \hline
                       & Avg. \#  \Algssf   & Precision & Coverage$^*$ & Avg. \# \Algextinpsource & Precision & Coverage$^*$\\ \hline
  CWE-78               & 6.12          & 85.42\%        &    100\%           &  4.72       & 76.17\%       &    100\%                \\
  CWE-134              & 6.71           & 86.2\%       &      100\%          &  4.52       & 80.21\%        &    100\%                \\
  CWE-190              & 5.2         & 90.21\%         &      100\%          &    5.63      & 72.14\%        &    100\%                      \\
  CWE-606              & 6.79         & 85.18\%        &     100\%          & 4.46     & 81.11\%           &    100\%            \\ \hline
  \end{tabular}
  \begin{tablenotes}
    \footnotesize
     \item[\quad \quad $^*$ the proportion of identification results covering the test subject target scenario.]
  \end{tablenotes}
\end{threeparttable}
  \end{table}

\finding{
\proj automatically identifies the sink and the source with an average precision of 82.1\%.
The identification results correctly
cover the expected sink and source pairs 
of the test subjects 100\%.
}

\subsection{Dangerous Flow Extraction (RQ3)}\label{sec:dangerous}

The effectiveness of \Algdangflow extraction
 affects the accuracy of \gptfour vulnerability checking.
 The FN of DFs directly leads to the FN of \proj vulnerability checks.
To evaluate the effectiveness of \Algdangflow extraction,
we analyzed vulnerable test subjects (labeled as "bad") in Juliet.
Vulnerable test subjects contain a potential vulnerability.
We check whether the DF extraction is successful
and whether the vulnerability triggering path between source and sink can be covered.

\begin{figure}[!htb]
	\centering
	\includegraphics[width=0.7\linewidth]{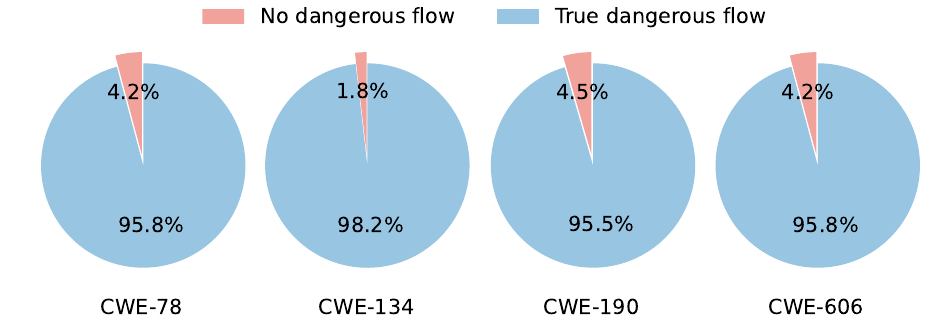}
	\caption{Dangerous flow extraction results.}
	\label{fig-dangerous}
\end{figure}

\cref{fig-dangerous} shows that among different types of vulnerable test subjects,
\proj successfully extracted the \Algdangflow for more than 95\% of the vulnerable test subjects.
Moreover, the \Algdangflow of each successfully extracted test subject is correct,
i.e. covering the code snippets
between source and sink that causes the test subject to trigger the vulnerability.
The detailed numbers of DFs corresponding to each CWE type are 920, 1178, 2731, and 230, respectively.
Each successfully extracted test subject corresponds to a unique DF.
This indicates that \proj's DF extraction is effective,
as \proj correctly captures potential paths
leading to vulnerabilities in the most of vulnerable test subjects.

However, there are test subjects where \proj did not successfully extract \Algdangflows (false negatives).
We investigated these test subjects and identified two primary reasons:
\circled{1} Ghidra decompilation error.
The decompilation engine may have problems parsing function call arguments,
such as three arguments being mistaken for two.
This leads to the breakage of taint data dependencies in the callee function,
ultimately affecting the extraction of \Algdangflows.
\circled{2} Pure virtual function calls.
Pure virtual functions have no function body and must be implemented in derived classes.
Test subjects with pure virtual function calls prevent \proj from extracting the \Algdangflow
because there is no concrete implementation to analyze.

\finding{\proj successfully and correctly extracts \Algdangflows covering more than 95\%
of test subjects containing potential vulnerabilities.
This demonstrates the ability of \proj to effectively extract \Algdangflows.
}

\subsection{Prompt Sequence Construction (RQ4)}
To evaluate the effectiveness of \AlgPromptSeq construction,
we designed five sets of comparative experiments based on the factors
that affect prompt construction: code slicing granularity and analysis task description granularity.
Specifically, keeping the same analysis task description as \proj,
we perform prompt construction on \circled{1} the entire program and \circled{2} the entire \Algdangflow respectively.
Moreover, we construct \AlgPromptSeqs based on more fine-grained slicing,
that is, \circled{3}
only retaining the code lines
related to the data dependency of the \Algssf
trigger argument.
For analysis task description,
we use functions in \Algdangflow as units
to construct \AlgPromptSeq consistent with \proj,
but we \circled{4} cancel the persona assignment
(\ie, the first prompt in \AlgPromptSeq
does not indicate that
the conversation will be conducted
in the role of a program analyst)
and \circled{5} do not split tasks
(\ie, directly instruct the GPT-4.0 to perform vulnerability analysis
without mentioning subtasks such as taint analysis and data dependency analysis).

\begin{figure}[!htb]
	\centering
	\includegraphics[width=\linewidth]{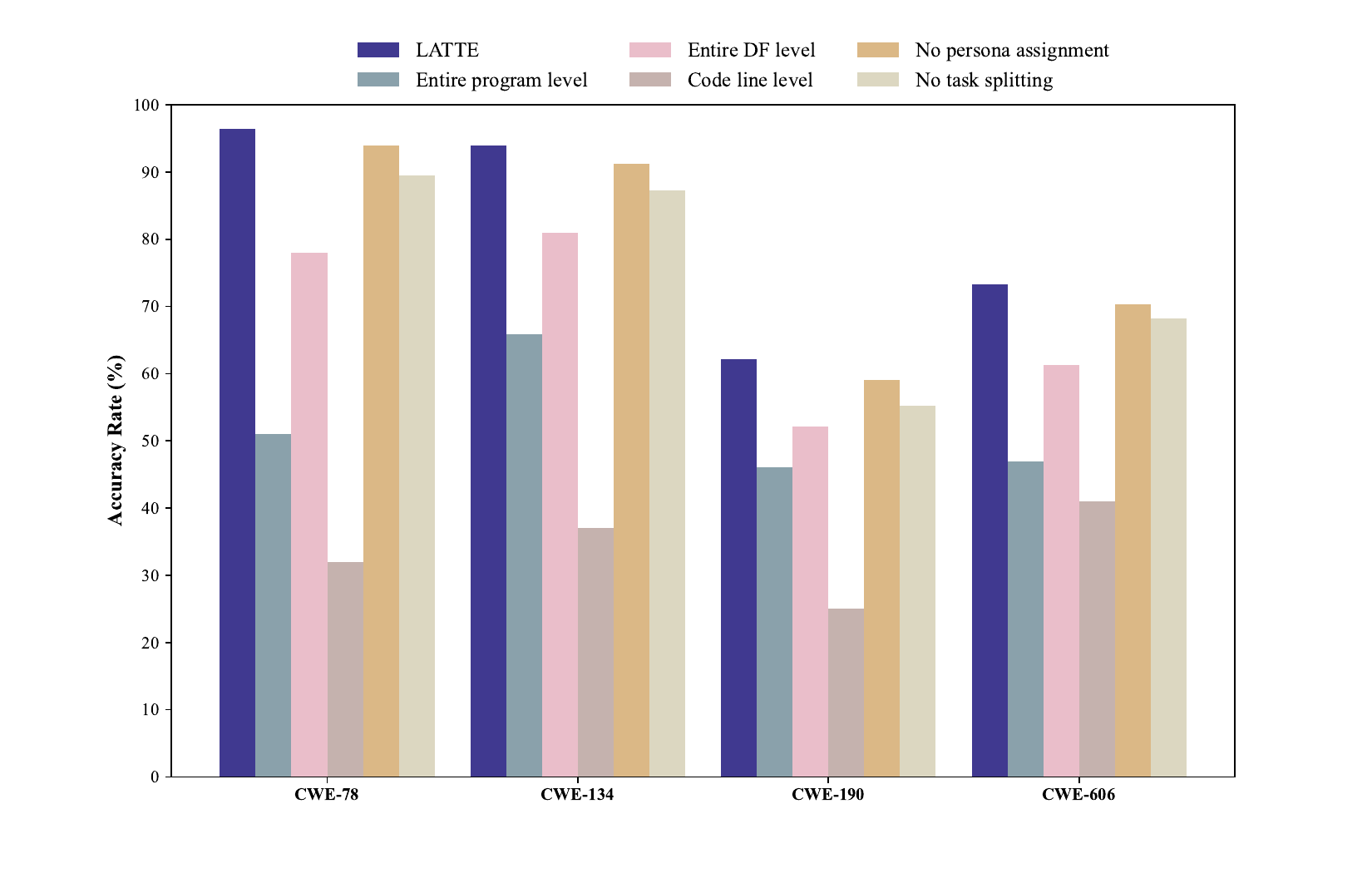}
	\caption{Comparison of the impact of PS construction on vulnerability analysis.}
	\label{fig-prompteval}
\end{figure}

Based on the above principles, we construct corresponding prompts (or prompt sequences)
for vulnerability analysis on the test subjects in
\cref{tab:juliet-eval}.
As shown in
\cref{fig-prompteval}, \proj outperforms the other five methods.
Moreover, 
the p-values for 
the comparison pairs 
calculated using the Mann-Whitney U Test~\cite{McNemar} 
shown in 
\cref{tab:dfpvalue}
are all less than 
\mycode{0.05}, 
indicating that the comparison  results 
are statistically significant.
Specifically,
\proj outperforms the approach on larger-grained slices (the entire program and the entire \Algdangflow)
and smaller-grained slices (code lines involved in tainted data) by an average accuracy of 28.7\%, 13.3\% and 46.4\%, respectively.
In large-granularity slices,
the entire \Algdangflow is also better than the entire program with an average accuracy of 15.4\%,
which shows that directly providing a large number of code snippets at one time
leads to a decrease in analysis accuracy.
For smaller-grained slicing,
\gptfour produced responses
that vulnerability analysis cannot be performed
due to lack of sufficient context.
This is due to the lack of associated code snippets
that prevent syntax analysis such as data flow and control flow.
Regarding the task description granularity,
\proj performs better than the method without persona assignment and task splitting,
with an average accuracy improvement of 2.78\% and 6.4\% respectively.
Compared to the granularity of code slicing,
changes in the task description have a relatively small impact
on the accuracy of the results.
The main reason is that the \llm itself has
sufficient understanding capability to reason and understand according
to the task requirements,
so persona assignment or task splitting does not have a significant impact on the results.
However, providing richer context is still helpful for task understanding.

\begin{table}[!htb]
	\centering
	\caption{The p-values calculated by Mann-Whitney U Test to illustrate the statistical significance of LATTE and comparative sets.}
	\label{tab:dfpvalue}
  \begin{tabular}{@{}llllll@{}}
			\hline
           & Entire program   &   Entire DF &  Code line  &  \begin{tabular}[c]{@{}c@{}}  No persona\\ assignment \end{tabular} & \begin{tabular}[c]{@{}c@{}} No task \\ splitting \end{tabular} \\ \hline
			CWE-78 &0.008 & 0.008 &0.008 & 0.008 &  0.012 \\ 
			CWE-134 & 0.008& 0.008 &0.008 & 0.012 &  0.012\\ 
			CWE-190 & 0.008 & 0.008 &0.008 &0.008 &  0.012 \\ 
			CWE-606 &0.008   &0.008 &0.008 & 0.012 &  0.012      \\ \hline

		\end{tabular}
\end{table}

\finding{Compared with prompts (prompt sequences)
constructed based on different code slicing granularities
and analysis task description granularities,
\proj demonstrates higher accuracy in detecting binary vulnerabilities.
}

\subsection{Real-world Vulnerability Inspection (RQ5)}\label{realsec}

RQ1-RQ4 evaluations demonstrated
the vulnerability analysis capabilities
of \proj on standard datasets.
To further illustrate \proj's
ability
to analyze unknown and real-world data,
we evaluated \proj
on 49 embedded device firmware samples
from \karonte.
LATTE issued a total of 255 alerts for 3,678 DFs.
As shown in \cref{tab:my-karonte}, 
the 255 alerts found after verification include 222 \tps.
of which there are 119 unique CWE-78 and CWE-120 bugs.

\begin{figure}[!htb]
    \centering
    \includegraphics[width=0.6\linewidth]{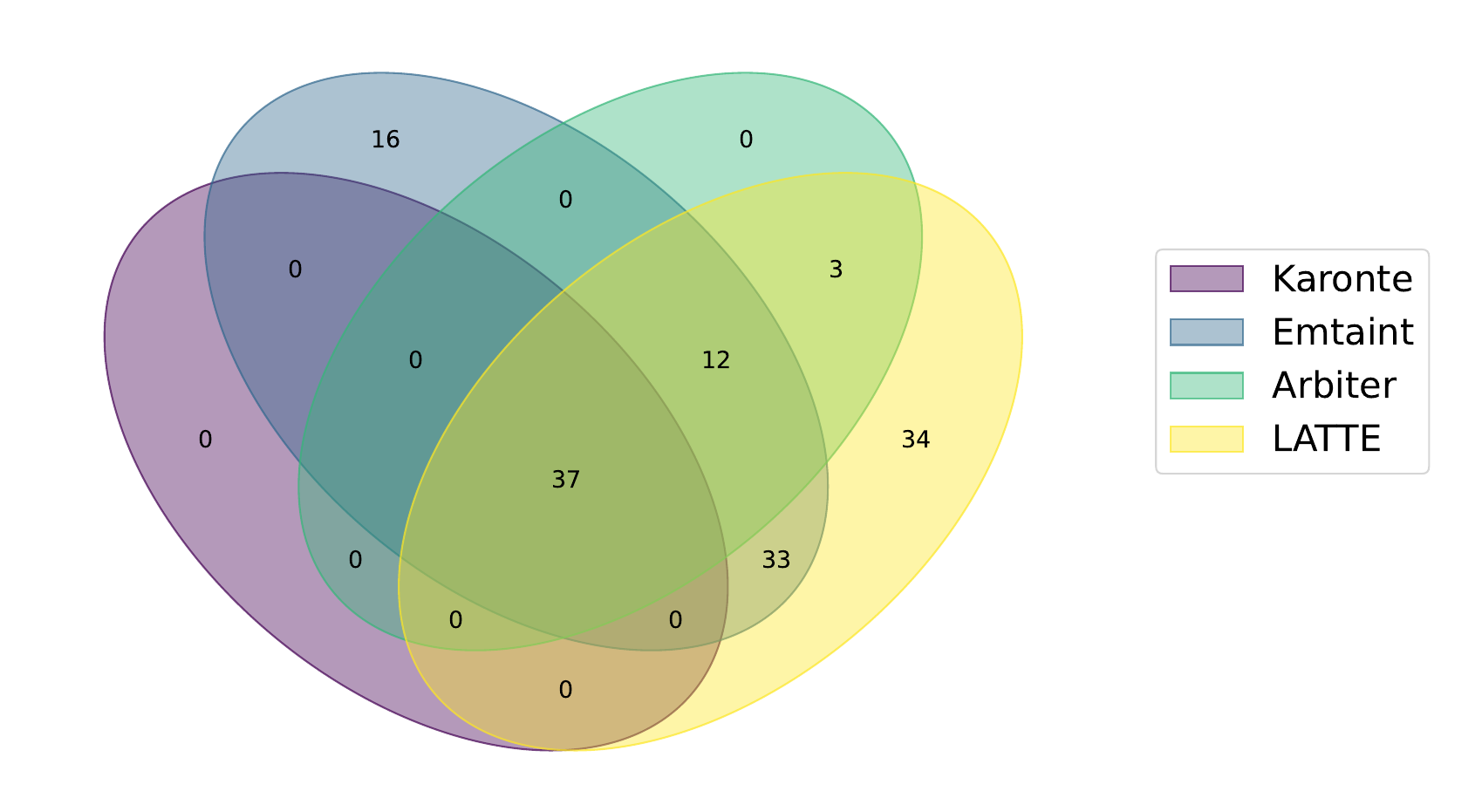}
    \caption{Unique bugs discovered by different tools.
    }
    \label{fig-unique}
  \end{figure}

\proj outperforms advanced binary taint analysis techniques.
\karonte and \arbiter found 46 and 151 \tps,
respectively,
while \proj found 222 \tps.
Moreover,
\proj 's precision 87.1\% is higher
than \karonte's 62.16\% and \arbiter's 48.67\%.
For analysis time, \karonte and \arbiter
rely on the
heavyweight technique
symbolic execution,
so \proj spends 425 and 109 minutes less time on analysis, respectively.
Although \emtaint
found 1296 more \tps than \proj,
many of \tps were the same bugs
caused by different
data flows.
much workload
to analysts to verify the results.
As shown in \cref{fig-unique},
the unique bugs found by \proj
cover the results of  \karonte and \arbiter,
while 16 unique bugs found by \emtaint were not found by \proj.
  The reason is that the pointer alias in \proj
  is not accurate,
  resulting in no relationship between variables pointing to the same memory.
  Therefore, the call chain extracted by \proj is incomplete,
  causing the corresponding \Algdangflow to be missed.
However, \proj found 119 unique bugs, 21 more than \emtaint.
Moreover, \proj found 37 previously unknown bugs,
and \cvebugs CVE numbers have been obtained due to the high threat.

\begin{table*}[htb]
  \caption{Vulnerability inspection results based on real-world firmware dataset collected by Karonte.}
  \label{tab:my-karonte}
  \begin{threeparttable}
  \resizebox{\textwidth}{!}{%
  \begin{tabular}{@{}llllllllllllllllllllll@{}}
  \hline
  \multirow{2}{*}{\textbf{Vendor}} & \multirow{2}{*}{\textbf{Samples}} & \multicolumn{5}{c}{\textbf{Karonte}}                                                                                                                                                                                                                                            & \multicolumn{5}{c}{\textbf{Emtaint}}                                                                                                                                                                                                                                            & \multicolumn{5}{c}{\textbf{Arbiter}}                                                                                                                                                                                                                                            & \multicolumn{5}{c}{\cellcolor[HTML]{C0C0C0}\textbf{\proj}}                                                                                                                                                                                                                                              \\ \cline{3-22}
                                   &                                   & \multicolumn{1}{c}{\textbf{Alerts}} & \multicolumn{1}{c}{\textbf{TP}} & \multicolumn{1}{c}{\textbf{FP}} & \multicolumn{1}{c}{\textbf{\begin{tabular}[c]{@{}c@{}}Unique\\ Bugs\end{tabular}}} & \multicolumn{1}{c}{\textbf{\begin{tabular}[c]{@{}c@{}}Avg.\\ Time\end{tabular}}} & \multicolumn{1}{c}{\textbf{Alerts}} & \multicolumn{1}{c}{\textbf{TP}} & \multicolumn{1}{c}{\textbf{FP}} & \multicolumn{1}{c}{\textbf{\begin{tabular}[c]{@{}c@{}}Unique\\ Bugs\end{tabular}}} & \multicolumn{1}{c}{\textbf{\begin{tabular}[c]{@{}c@{}}Avg.\\ Time\end{tabular}}} & \multicolumn{1}{c}{\textbf{Alerts}} & \multicolumn{1}{c}{\textbf{TP}} & \multicolumn{1}{c}{\textbf{FP}} & \multicolumn{1}{c}{\textbf{\begin{tabular}[c]{@{}c@{}}Unique\\ Bugs\end{tabular}}} & \multicolumn{1}{c}{\textbf{\begin{tabular}[c]{@{}c@{}}Avg.\\ Time\end{tabular}}} & \multicolumn{1}{c}{\textbf{Alerts}} & \multicolumn{1}{c}{\textbf{TP}} & \multicolumn{1}{c}{\textbf{FP}} & \multicolumn{1}{c}{\textbf{\begin{tabular}[c]{@{}c@{}}Unique\\ Bugs\end{tabular}}} & \multicolumn{1}{c}{\textbf{\begin{tabular}[c]{@{}c@{}}Avg.\\ Time\end{tabular}}} \\ \hline
  NETGEAR                          & 17                                & 36                                  & 23                              & 13                              &   19                                                                                & 17:13                                                                            & 849                                 & 849                             & 0                               &   47                                                                                & 00:05                                                                            &  131                                   &    63                             &     68                            &    26                                                                                &          02:20                                                                        &   142                                 &   124                              &     18                            &   54                                                                                 & 00:38                                                                                \\
  D-Link                           & 9                                 & 24                                  & 15                              & 9                               &   10                                                                                & 14:09                                                                            & 299                                 & 234                             & 65                              &     29                                                                               & 00:02                                                                            &   99                                  &      48                           &       51                          &      14                                                                              &           03:18                                                                       &   65                                  &     59                            &     6                            &  42                                                                                  &      00:21                                                                            \\
  TP-Link                          & 16                                & 2                                   & 2                               & 0                               &   2                                                                                 & 01:30                                                                            & 73                                  & 73                              & 0                               &     3                                                                               & 00:05                                                                            &    21                                 &       18                          &         3                        &        6                                                                            &             02:40                                                                     &    29                                 &      23                           &     6                            &   10                                                                                 &       00:33                                                                          \\
  Tenda                            & 7                                 & 12                                  & 6                               & 6                               &   6                                                                                & 01:01                                                                            & 362                                 & 362                             & 0                               &     19                                                                               & 00:05                                                                            &    49                                 &      22                           &       27                          &      6                                                                              &             03:25                                                                     &   19                                  &     16                            &    3                             &  13                                                                                  &      00:29                                                                            \\ \hline
  Total                            & 49                                & 74                                  & 46                              & 28                              &   37                                                                                 & 451:06                                                                           & 1583                                & 1518                            & 65                              &    98                                                                                & 03:38                                                                            &   300                                  &    151                             &    149                            &   52                                                                                 &         135:57                                                                         &  255                                    &   222                             &    33                             &  119                                                                                  &    26:06                                                                              \\ \hline
  \end{tabular}%
  }
  \begin{tablenotes}
    \footnotesize
    \item[\textbf{Alerts:} the number of bugs considered by the technique. ]
    \item[\textbf{TP:} the vulnerability trigger location is reachable by user-controllable data.]
    \item[\textbf{FP:} the vulnerability trigger location is unreachable or not caused by user-controllable data.]
    \item[\textbf{Unique Bugs:} different pairs of user-controllable data receiving locations and vulnerability triggering locations in TPs.]
    \item[\textbf{Avg. Time:} average time spent analyzing a firmware represented by \mycode{hh:mm}.]
  \end{tablenotes}
\end{threeparttable}
  \end{table*}

  \finding{\proj
  effectively found
  \totaluniquebugs unique bugs (including \newbugs previously unknown bugs)
  in the real-world firmware dataset,
  outperforming the state-of-the-art techniques \karonte, \emtaint, and \arbiter.
  }

\section{Discussion}

\subsection{Limitations on Vulnerability Types}
\proj faces challenges in handling certain types of vulnerabilities,
such as logic vulnerabilitie~\cite{CWE840},
race condition vulnerabilities~\cite{CWE362},
and vulnerabilities involving complex mathematical calculations~\cite{10.1145/2664243.2664282,CWE310}.
These vulnerabilities often involve intricate code snippets
with complex nesting and jumping,
making them difficult for the \llm to effectively analyze.
Therefore, the under/over-tainting problem is prone to occur.
Moreover, these vulnerabilities
are also difficult to detect
by using taint analysis alone.
Complex vulnerabilities
require the extraction of more
context information
such as resource locks,
interrupt vector tables,
shared memory addresses, \etc
to assist in analysis.
Collecting more valuable context information and
providing it to \llms for vulnerability analysis
is a direction worthy of further research.

\subsection{Limitations on Binary Analysis}\label{subseclimi}
\proj implements binary analysis based on Ghidra~\cite{ghidra},
including reverse engineering, function structure recovery, and calling relationship analysis.
Ghidra provides relatively complete analysis algorithms and APIs for developers to use.

However, binary analysis itself is inherently complex and poses several challenges~\cite{meng2016binary}.
Many issues remain partially or entirely unsolved.
For example, Ghidra’s analysis can struggle to determine when two pointers refer to the same memory location,
leading to potential inaccuracies in data flow tracking and alias analysis.
Additionally, Ghidra sometimes has difficulty resolving virtual calls,
which can obscure the true calling relationships in the binary.
This limitation is especially problematic in C++ programs with extensive use of polymorphism,
where the inability to accurately resolve virtual calls
can hinder the understanding of control flow and data dependencies.
LLM excels at identifying patterns in code and data flow, and it would be 
worth exploring further in the future to assist in identifying common indirect control flow patterns.

Moreover, when the compiler inlines a function,
the original boundary between the caller function
and the callee function is eliminated,
making it difficult to reconstruct the original function structure
and understand the precise calling relationship.
Fortunately,
if a callee is inlined,
the caller becomes larger by incorporating the implementation of the callee.
Therefore, it does not affect the analysis
of the code with \llms.
Moreover,
function inlining affects
the identification of statically-linked source and sink functions.
But in most cases,
the source and sink are provided by the system library~\cite{vadayath2022arbiter,gotovchits2018saluki}.
In order to ensure the size and compatibility of the program,
the system library is usually
dynamically linked,
\eg, glibc~\cite{vadayath2022arbiter,redini2020karonte}.

\subsection{Importance of Metrics for Evaluating Vulnerability Analysis}

Precision and recall are two key metrics
for real program vulnerability analysis.
Analytical methods often require
a balance between false positives (\fps)
and false negatives (\fnns)~\cite{vadayath2022arbiter}.
Precise analysis algorithms can reduce false positives,
but may overlook diverse potential vulnerabilities,
thus increasing false negatives.
Conversely, over-approximation analysis algorithms may identify
more potential vulnerabilities, but also increase the number of false positives.
Therefore, security engineers need to choose the appropriate method
based on specific analysis contexts and objectives.
Furthermore, identifying unique bugs is crucial for
both precision and recall metrics.
Duplicate true positives (\tps)
not only add no value to security analysis
but also increase the overhead involved
in validating vulnerabilities~\cite{redini2020karonte,cheng2023emtaint}.

The following guidelines provide tool selection recommendations
based on analysis costs, business scenarios, and security requirement standards~\cite{mcgraw2006software,nist80053,iso27001}.
The recall score is particularly important
when resources such as time and labor are sufficient.
Additionally, using high-recall tools to find as many vulnerabilities
as possible is especially useful in the early stages or when reviewing new code.
On the other hand, when dealing with a large number of subjects
and limited resources,
security engineers should prioritize
high-precision tools
to ensure that reported vulnerabilities are genuine,
thus minimizing the time spent on false positives.
For systems requiring stringent security measures,
use a combination of tools to balance precision and recall,
ensuring comprehensive
coverage and accurate identification of vulnerabilities.

According to the evaluation results in~\cref{subsec:effect},
\proj outperforms other tools in both of precision and recall.
Meanwhile, \proj found more unique bugs in the evaluation of real firmwares 
in~\cref{realsec},
and covered the bug results of Arbiter and Karonte.
Based on the above guidelines, engineers should give priority to LATTE. 
However, the cost of \llm tokens for using \proj is unavoidable.
Therefore, if engineers have sufficient token resources, 
\proj is recommended to be used over traditional methods.

\subsection{Impact of Binary Protection on \proj}
\proj performs binary analysis based on the intermediate representation P-Code
of Ghidra,
enabling binary analysis across different architectures
such as X86, X64, ARM, and MIPS.
 Consistent with other binary vulnerability analysis techniques~\cite{redini2020karonte,cheng2023emtaint,chen2021sharing,vadayath2022arbiter},
 \proj cannot directly analyze binaries that have been obfuscated or encrypted.
 Additional steps are needed to ensure the binary can be processed by Ghidra.
 There is ongoing research focused on deobfuscation and decryption,
 which can aid subsequent vulnerability analysis~\cite{madou2006loco,vasselle2019breaking,tofighi2019defeating,garba2019saturn}.

 Moreover,
 the performance of \proj
 was evaluated on binaries lacking debugging information,
 symbol tables, and comments.
 Maliciously tampered semantics can interfere with \proj's analysis.
 For example, renaming a function containing a vulnerability to “no\_vulnerability” might increase
 the likelihood of misjudgment.
 However, by instructing \llms to ignore such semantic information, the impact can be significantly reduced.
 Offensive and defensive techniques in binary analysis remain an ongoing challenge,
 and our findings contribute to understanding the capabilities and limitations of LLMs in this context.

\subsection{Identification of Sources and Sinks}
The specification of sources and sinks is necessary to initiate \proj analysis.
If the analyst already has the identified sources and sinks, 
the source and sink identification stage can be skipped and 
DF extraction and PS construction 
can be performed directly by setting the sources and sinks.
If not, 
the analyst can use \proj to help identify sources and sinks. 
We evaluated source and sink identification based on the
Juliet test suite in~\cref{source-sink-iden}.
The identification results of sources and sinks can be reused
if another binary uses the same dynamic library.
We conducted vulnerability inspections
using these identification results in realwolrd firmwares in
\cref{realsec}.

In the real world, identifying sources and sinks is a complex problem.
First, real programs require more queries 
due to the presence of various linked libraries.
For example, the glibc requires about 3000-4000 queries, 
and the OpenGL requires several hundred queries.
Therefore, even with \proj to assist 
in identifying sources and sinks for caching, 
it requires a certain amount of time overhead.
Second, as described in~\cref{subseclimi}, 
although most binary prorams use dynamic linking, 
static linking and inlining make it more difficult to identify sources and sinks.
\proj needs to be combined with program analysis techniques such as 
code summarization and dynamic analysis to further determine the source and sink.
Third,
some custom functions in real-world binaries 
may also be used as tainted sources and sinks~\cite{liufits}, 
which further complicates the identification of sources and sinks.
\llms can recognize well-known functions like
\mycode{recv} and \mycode{printf},
its performance is limited when encountering previously unseen functions.
Meanwhile, the hallucination of \llms seriously 
affects the identification of sources and sinks.
Continuously refining and optimizing prompts or models based on 
expert feedback, code information, 
and specification data is critical 
to meeting the challenges of various libraries and custom functions.

\subsection{Applicability to Source Code}
While \proj is designed for vulnerability analysis of binary code,
its underlying concept is adaptable for source code analysis as well. 
If source code analysis is required, 
three components of the \proj workflow (outlined in~\cref{fig-overview}) 
would need adjustment.
First, source code analysis bypasses the need for disassembly or decompilation,
allowing for direct code analysis to identify dangerous flows. 
Second, the algorithm for extracting dangerous flows 
must be modified to work with source code.
For example, in decompiled code, 
the analysis relies on registers and memory addresses, 
whereas, in source code, it focuses on variable analysis. 
Furthermore, binary code typically integrates 
all calling relationships and dependencies into a single file, 
while source code requires parsing dependencies across multiple files. 
Third, the prompt templates need to be optimized. 
The start template in binary analysis states that the code 
is "generated by decompilation," 
which provides helpful context. 
When analyzing source code, 
removing this statement or 
indicating that the code is source code 
can help LLMs analyze the source code more directly, 
benefiting from explicit variable names and comments.
Providing additional descriptive information further 
aids LLMs in analyzing the code effectively.

To demonstrate the feasibility of the concept of \proj  for source code, 
we utilized SVF~\cite{svf} 
to extract dangerous flows from the source code of the Juliet test cases 
(\cref{sec:evaluation}) 
and removed the “generated by decompilation” statement 
from the start prompt. 
As shown in \cref{tab:sourcecompare}, 
the accuracy of detecting 
vulnerabilities in source code for CWE-78, CWE-134, CWE-190, 
and CWE-606 improved by 1.64\%, 2.91\%, 17.9\%, and 5.07\% respectively, 
compared to stripped binary analysis. 
The p-values for these comparisons, 
calculated using the Mann-Whitney U Test,
were all less than \mycode{0.1}, 
indicating statistical significance. 
This improvement results from the semantic clarity in source code, 
with function names, variable names, 
and type definitions providing critical 
context that aids LLMs 
in understanding and analyzing 
vulnerabilities.

However, when combined with LLMs, 
there may be more effective approaches than LATTE for source code analysis. 
For instance, integrating LLMs with advanced program analysis 
tools such as CodeQL~\cite{he2023large,surameery2023use,wadhwa2023frustrated} 
could provide a more comprehensive understanding of 
code structure and semantics. 
Additionally, pointer analysis and other techniques 
could be applied directly to source code, rather than decompiled code, 
to further enhance the accuracy and effectiveness of vulnerability detection.

\begin{table}[t]
	\centering
	\caption{Comparison of vulnerability checking accuracy on source code and decompiled code.}
	\label{tab:sourcecompare}
	\begin{tabular}{@{}lllll@{}}
	\hline
	         & Stripped Binary &Source Code & Improvement & p-value\\ \hline
	CWE-78   &96.46\%  &98.1\%        &       1.64\%     & 0.095 \\
	CWE-134  & 93.88\%&96.9\%          &      3.02\%     & 0.008 \\
	CWE-190  & 62.1\% &80\%             &      17.9\%    & 0.008  \\
	CWE-606  & 73.33\% &78.4\%           &     5.07\%    & 0.004   \\ \hline
	\end{tabular}%
	\end{table}
\section{Related Work}

\myparagraph{Taint Analysis}
Taint analysis is a software security analysis method
that aims to identify and analyze potential security vulnerabilities in programs.
It is widely used for analysis in source code, web applications and Android programs.
VulSlicer~\cite{SALIMI2022111450} and FICS~\cite{263838}
performed data flow analysis on the source and sink pairs in the source code.
Merlin~\cite{10.1145/1543135.1542485}, Julian Thomé et al.~\cite{THOME2018766}
and JOZA~\cite{naderi2015joza}
summarized artificial experience into rules and used heuristic methods
to discover taint-style vulnerability in web applications.
DroidSafe~\cite{gordon2015information}, SCANDAL~\cite{kim2012scandal} and AndroidLeaks~\cite{gibler2012androidleaks}
identified some API calls or callback functions as taint sources
in Android programs to perform taint analysis.
The main difference between taint analysis of binary programs and the above
program types is the loss of information such as program semantics and code structure.
Therefore, when implementing a binary taint analysis engine, more expert experience
and more manual work are required.
Specific binary taint analysis techniques are introduced in the next paragraph.

\myparagraph{Binary Vulnerbility Analysis}
For vulnerability analysis of binary,
there are both dynamic and static techniques.
Dynamic techniques monitor unexpected behaviors of test targets as they execute test cases,
such as fuzzing~\cite{feng2021snipuzz,10.1145uafl,scharnowski2022fuzzware,chen2018iotfuzzer,zheng2019firm},
concolic execution~\cite{stephens2016driller,poeplau2020symbolic,yun2018qsym,coppa2022symfusion}, emulation analysis~\cite{kim2020firmae,hernandez2022firmwire,clements2020halucinator,mera2021dice,feng2020p2im}.
The results of dynamic analysis are reliable,
but the test code coverage is low and the
test execution depends on the specific operating environment.
Binary execution environments vary depending on the operating system,
architecture, and even peripherals.
As a result, the application and performance of dynamic analysis is often limited.
Static analysis techniques are suitable for binary analysis
due to their high code test coverage and not limited to the execution environment.
BootStomp~\cite{redini2017bootstomp}, Saluki~\cite{gotovchits2018saluki}, Dtaint~\cite{cheng2018dtaint}, Karonte~\cite{redini2020karonte}, SATC~\cite{chen2021sharing},
Arbiter~\cite{vadayath2022arbiter} and \emtaint~\cite{cheng2023emtaint} techniques
primarily leverage taint analysis
combined with program analysis, symbol execution, and machine learning
to analyze binary vulnerabilities.
However, these methods commonly rely on human experience to define propagation and inspection rules.
\proj, on the other hand, is the first to propose a taint analysis technique
combined with \llms for binary vulnerability inspection,
offering a novel approach that avoids the need for manual rule definition.

\myparagraph{LLM-assisted Code Analysis}
The \llm is a class of deep learning models
that have significantly impacted natural language processing.
In recent years, with the excellent code understanding ability of \llms,
they have been widely applied in code-related tasks,
such as code generation~\cite{pearce2022asleep,ahmad2021unified,fried2022incoder,yetistiren2022assessing},
patch generation~\cite{xia2022less,xia2023conversational,sobania2023analysis,prenner2022can},
and code summarization~\cite{ahmed2022few}.
For code vulnerability analysis,
it is often necessary to combine various program analysis techniques
on the basis of \llms' code understanding to conduct effective vulnerability auditing.
For instance, FuzzGPT~\cite{deng2023fuzzgpt} and TitanFuzz~\cite{deng2023large} utilize the \llm to analyze existing code snippets and generate test cases to assist in API fuzzing.
CODAMOSA~\cite{lemieux2023codamosa} uses the \llm to analyze code coverage information and modify test cases to improve test coverage.
Moreover, 
LLMDFA~\cite{wangllmdfa}, inspired by summary-based data flow analysis, introduced the dadaflow analysis framework supported 
by \llm and explored the idea of \llm-assisted source code analysis. 
LLMSAN~\cite{wangsanitizing} further used the ideas of few-shot chain-of-thought and program properties 
to reduce the impact of \llm hallucination on the reliability of data flow analysis.
Unlike these \llm-assisted analysis techniques,
\proj utilizes the \llm to assist static analysis and works on binaries that provide less code information than source code.

\myparagraph{Security Slicing}
Obtaining code snippets from a program for vulnerability analysis is a popular security analysis method. To reduce the time overhead of verifying warnings and improve inspection accuracy, the results of code slicing are often combined with other analysis and testing techniques for further verification.
FuzzSlice~\cite{FuzzSlice} and Helium~\cite{Helium}
used static analysis technology
to construct compilable harnesses from source code slices,
 which are then subjected to vulnerability verification through fuzzing.
 SFuzz~\cite{SFuzz} implemented fuzzing on independent code snippets
 extracted from real-time operating systems binaries.
 These techniques must ensure that the code snippets
 not only contain the entire input processing flow
 but also that the slicing result is compilable and executable.
 To address the constraints on the execution conditions of code snippets,
 Engler et al.~\cite{UC-KLEE}
 proposed under-constrained symbolic execution,
 allowing symbolic execution of any function
 without initializing any of its data structures or setting up any environment.
 However, this technique still faces challenges with loops, pointers
 and library function calls, affecting analysis accuracy.
 Our approach, \proj, leverages the idea of taint analysis to slice the program
 and then utilizes \llms for vulnerability analysis.
 By using \llms, our approach reduces the dependency on the execution environment,
 alleviating the need for comprehensive environment setup and initialization.
\llms provide a deeper semantic understanding of the code, improving the identification and analysis of vulnerabilities within the sliced snippets.
 Combining taint analysis with \llms
 allows for a more focused and relevant slicing,
  ensuring that the critical parts of the code related
  to potential vulnerabilities are included in the analysis.
\section{Conclusion}

This paper proposes \proj, the first \llm-powered static binary taint analysis.
Traditional taint analyzers rely on human expertise to manually define patterns and rules,
and this manual nature severely limits the effectiveness and performance of
the taint analysis technique.
\proj performs dangerous flow slicing by slicing vulnerable destinations back to external input sources.
\proj then combines code snippets to construct prompt sequences, leveraging \llms'
context-aware assistance and human-like understanding capabilities for code vulnerability inspection.
Based on the substantial evaluations,
\proj outperforms the state of the art across all taint-style vulnerability types,
such as buffer overflow and command injection.
\proj sheds light on the combination of program analysis and \llms for vulnerability analysis.

\section{Acknowledgements}
We thank the anonymous reviewers for their valuable comments to improve our paper. 
This work was supported by Ant Group Postdoctoral Programme.

\bibliographystyle{ACM-Reference-Format}
\bibliography{ccs-sample}

\end{document}